\documentclass[twocolumn]{aastex62}
\pdfoutput=1 
\usepackage{amsmath,amstext}
\usepackage[T1]{fontenc}
\usepackage[figure,figure*]{hypcap}
\usepackage{mwe}
\usepackage{lipsum}
\usepackage{graphicx} 
\usepackage{multirow}
\usepackage{xcolor,colortbl}
\usepackage{xspace}
\usepackage{csvsimple}
\usepackage{afterpage}

\newcommand{\twelveCO}{$^{12}$CO}
\newcommand{\thirtCO}{$^{13}$CO}

\newcommand{\nh}{$n_{\text{H}_2}$\xspace}

\newcommand{\XCO}{$X_\text{CO}$\xspace}
\newcommand{\XthCO}{$X_\text{13CO}$\xspace}
\newcommand{\radex}{\texttt{RADEX}\xspace}
\newcommand{\alphavir}{$\alpha_\text{vir}$\xspace}

\shorttitle{LMC Ridge Structural Analysis}
\shortauthors{Finn et al.}

\begin{document}

\title{Structural and Dynamical Analysis of the Quiescent Molecular Ridge in the Large Magellanic Cloud}

\author[0000-0001-9338-2594]{Molly K. Finn} %
\affiliation{Department of Astronomy, University of Virginia, Charlottesville, VA 22904, USA}

\author[0000-0002-4663-6827]{Remy Indebetouw} %
\affiliation{Department of Astronomy, University of Virginia, Charlottesville, VA 22904, USA}
\affiliation{National Radio Astronomy Observatory, 520 Edgemont Road, Charlottesville, VA 22903, USA}

\author[0000-0001-8348-2671]{Kelsey E. Johnson} %
\affiliation{Department of Astronomy, University of Virginia, Charlottesville, VA 22904, USA}

\author[0000-0002-7408-7589]{Allison H. Costa} %
\affiliation{National Radio Astronomy Observatory, 520 Edgemont Road, Charlottesville, VA 22903, USA}

\author[0000-0002-3925-9365]{C.-H. Rosie Chen} %
\affiliation{Max-Planck-Institut f\"{u}r Radioastronomie, Auf dem H\"{u}gel 69, D-53121 Bonn, Germany}

\author[0000-0001-7813-0380]{Akiko Kawamura} %
\affiliation{National Astronomical Observatory of Japan, National Institutes of Natural Sciences, 2-21-1 Osawa, Mitaka, Tokyo 181-8588, Japan}

\author[0000-0001-7826-3837]{Toshikazu Onishi} %
\affiliation{Department of Physics, Graduate School of Science, Osaka Metropolitan University, 1-1 Gakuen-cho, Naka-ku, Sakai, Osaka 599-8531, Japan}

\author[0000-0001-8224-1956]{J\"{u}rgen Ott} %
\affiliation{National Radio Astronomy Observatory, P.O. Box O, Socorro, NM 87801, USA}
\affiliation{Physics Department, New Mexico Institute of Mining and Technology, 801 Leroy Pl., Socorro, NM 87801, USA}

\author[0000-0003-2248-6032]{Marta Sewi{\l}o} %
\affiliation{Exoplanets and Stellar Astrophysics Laboratory, NASA Goddard Space Flight Center, Greenbelt, MD 20771, USA}
\affiliation{Department of Astronomy, University of Maryland, College Park, MD 20742, USA}
\affiliation{Center for Research and Exploration in Space Science and Technology, NASA Goddard Space Flight Center, Greenbelt, MD 20771}

\author[0000-0002-2062-1600]{Kazuki Tokuda} %
\affiliation{Department of Earth and Planetary Sciences, Faculty of Sciences, Kyushu University, Nishi-ku, Fukuoka 819-0395, Japan}
\affiliation{National Astronomical Observatory of Japan, National Institutes of Natural Sciences, 2-21-1 Osawa, Mitaka, Tokyo 181-8588, Japan}

\author[0000-0002-7759-0585]{Tony Wong} %
\affiliation{Astronomy Department, University of Illinois, 1002 W. Green Street, Urbana, IL 61801, USA}

\author[0000-0001-6149-1278]{Sarolta Zahorecz} 
\affiliation{National Astronomical Observatory of Japan, National Institutes of Natural Sciences, 2-21-1 Osawa, Mitaka, Tokyo 181-8588, Japan}
\affiliation{Department of Physics, Graduate School of Science, Osaka Metropolitan University, 1-1 Gakuen-cho, Naka-ku, Sakai, Osaka 599-8531, Japan}

\begin{abstract}

We present a comparison of low-J \thirtCO\ and CS observations of four different regions in the LMC---the quiescent Molecular Ridge, 30~Doradus, N159, and N113, all at a resolution of $\sim3$ pc. The regions 30~Dor, N159, and N113 are actively forming massive stars, while the Molecular Ridge is forming almost no massive stars, despite its large reservoir of molecular gas and proximity to N159 and 30~Dor. We segment the emission from each region into hierarchical structures using dendrograms and analyze the sizes, masses, and linewidths of these structures. We find that the Ridge has significantly lower kinetic energy at a given size scale and also lower surface densities than the other regions, resulting in higher virial parameters. This suggests that the Ridge is not forming massive stars as actively as the other regions because it has less dense gas and not because collapse is suppressed by excess kinetic energy. We also find that these physical conditions and energy balance vary significantly within the Ridge and that this variation appears only weakly correlated with distance from sites of massive star formation such as R136 in 30~Dor, which is $\sim1$ kpc away. These variations also show only a weak correlation with local star formation activity within the clouds.

\end{abstract}


\section{Introduction} \label{sec:intro}

Star formation is a crucial component of our understanding of galactic environments. As we study galaxies at a wide variety of distance scales and size scales, we look for easily-observable tracers of star formation and molecular gas behavior, such as the Schmidt-Kennicutt law relating molecular gas surface density and galactic star formation rate \citep{Kennicutt98}. These relations have had success at predicting star formation rates in many environments, but we also know of cases where those relations do not hold. For example, the Central Molecular Zone of our own {Galaxy} is forming stars an order of magnitude slower than we would expect for a region with so much molecular gas available \citep{Longmore13}. 
It is clear that while these relations hold well on galactic scales, the physical conditions that give rise to such relations are not constant across different environments within a galaxy.

The Large Magellanic Cloud (LMC) is an excellent laboratory in which to study the variations in galactic environments. At a distance of 50 kpc \citep{Pietrzynski13}, it is close enough that we can resolve individual molecular clouds, while its low inclination angle \citep[$\sim34^\circ$; ][]{vanderMarel14} allows us a clear view of the contents of the galaxy with little line-of-sight or distance confusion. It also hosts a wide variety of star-forming environments, making comparisons between them relatively straightforward since they share a common distance, metallicity \citep[$Z\sim0.5Z_\odot$; ][]{Rolleston02}, and galactic environment. In this paper, we focus our analysis on the quiescent Molecular Ridge, 30~Doradus (30~Dor), N159, and N113 to span the extremes of star formation in the LMC. 

\begin{figure*}
    \centering
    \includegraphics[width=0.85\textwidth]{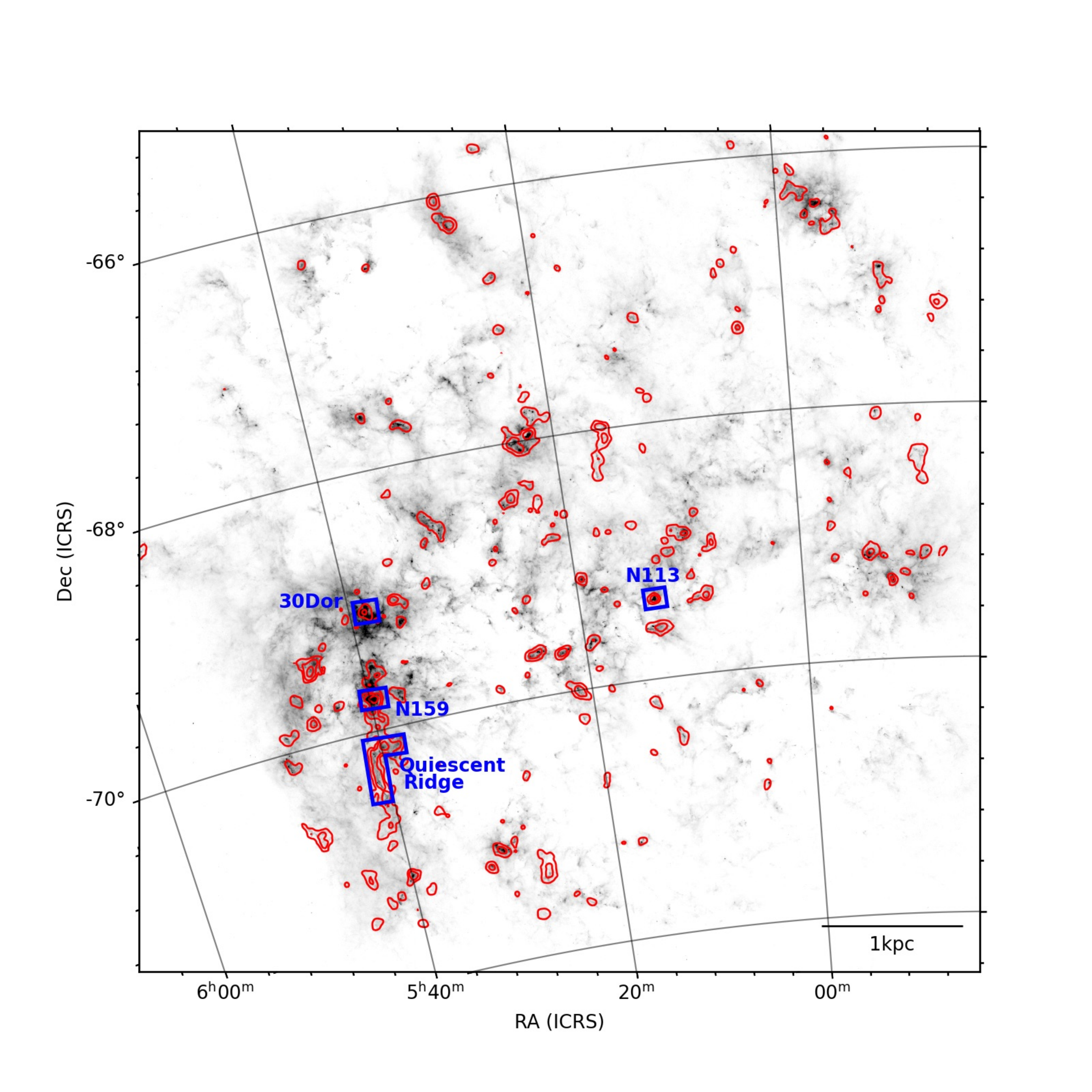}
    \caption{The LMC with the Molecular Ridge, 30~Dor, N159, and N113 highlighted in blue. The grayscale is PACS 250 $\mu$m from the HERITAGE survey \citep{Meixner13}, and the red contours are \twelveCO(1-0) from the NANTEN survey \citep{NANTEN} showing the extent of all the CO-bright molecular gas in the LMC. Levels are 1.6, 5, and 12 K km s$^{-1}$ }
    \label{fig:LMC map}
\end{figure*}

The Molecular Ridge (also referred to throughout this work as ``the Ridge'') is a $\sim2$ kpc-long structure in the LMC and contains nearly 30\% of all the CO-bright molecular gas mass in the galaxy \citep[][see Figure \ref{fig:LMC map}]{Cohen88,NANTEN,Mizuno01}. Despite this large quantity of molecular gas, the southern part of the Ridge is forming very few massive stars, based on low H$\alpha$ emission and lack of optically-identified young massive clusters \citep{Davies76, Bica96, Yamaguchi01}. 

\cite{Indebetouw08} found that the Schmidt-Kennicutt law \citep{Kennicutt98} predicts that the Ridge should have a star formation rate of $8\times10^{-3}$ M$_\odot$ yr$^{-1}$, but the total H$\alpha$ and 24$\mu$m emission in the Ridge suggests a star formation rate of only $2.6\times10^{-4}$ M$_\odot$ yr$^{-1}$ \citep{Calzetti07}. \cite{Indebetouw08} found that these numbers can be better reconciled by looking for embedded young stellar objects (YSOs) from the Spitzer Space Telescope survey Spitzer Surveying the Agents of Galaxy Evolution \citep[SAGE; ][]{Meixner06}, which brought the measured star formation rate to approximately $4\times10^{-3}$ M$_\odot$ yr$^{-1}$, within a factor of two of the value predicted by the Schmidt-Kennicutt law. This suggests that the lack of H$\alpha$ and optical clusters is because the Ridge is preferentially forming low-mass star clusters. 

This is a stark contrast to regions just north of the Ridge, such as 30~Dor and N159 (see Figure\,\ref{fig:LMC map}), which are some of the most active massive-star-forming regions in the LMC. 30~Dor is home to R136, the closest known super star cluster (SSC), and other young, massive clusters that bring the total mass of recently formed stars in the region up to $\sim8.7\times10^4$ M$_\odot$ \citep{Cignoni15}. 
There are still giant molecular clouds (GMCs) forming stars in the region, although the current star formation in 30~Dor is less extreme than it once was, forming primarily low- and intermediate-mass stars \citep{Walborn13,Sabbi16}. 
Just south of 30~Dor is N159, which contains several embedded high-mass YSOs and HII regions \citep{Chen10}, suggesting that massive star formation is ongoing in this region. N113 similarly has signs of active high-mass star formation \citep{Sewilo10,Seale12,Ward16}. By directly comparing the molecular gas in these regions with that in the Ridge, we hope to identify differences in the physical conditions that could suggest why the regions differ so much in star forming properties.

\cite{Finn21} fit \radex models to CO emission in the Ridge and found that the fitted volume density, \nh, had the strongest correlation with the presence of YSOs associated with the CO clumps. They hypothesized that the Ridge could be forming massive stars so sluggishly either because the molecular gas is lower density than the other star forming regions to its north, or because the threshold density for star formation is higher in the Ridge. For example, the latter could be caused by higher amounts of turbulent or magnetic support suppressing collapse in the Ridge. 

In this paper, we seek to understand the underlying differences in physical conditions between the Ridge and the other massive star forming regions including 30~Dor, N159, and N113. In \S\ref{sec:obs}, we present the different observations and regions being compared in this analysis, and in \S\ref{sec:dendrogram} we describe how we segment that observed emission into smaller structures using both dendrogram and clump-finding methods. We calculate the mass, velocity dispersion, and radii of these structures in \S\ref{sec:properties}. We compare the different regions by fitting size-linewidth relations in \S\ref{sec:SL plots} and by considering their virial balance of turbulent and gravitational energy in \S\ref{sec: virial plot}. We examine variations in these physical properties within the Ridge in \S\ref{sec: ridge variation} and look at the spatial dependence of those variations in \S\ref{sec: spatial dependence}. We bring the results of all of these sections together in \S\ref{sec:discussion} to discuss the overall picture of the differences in these regions, then summarize our conclusions in \S\ref{sec:conclusions}.

\section{Observations} \label{sec:obs}

We examine the Molecular Ridge in \thirtCO(1-0) and CS(2-1) and compare it to three other regions in the LMC: 30~Dor, N159, and N113. The observations used in this analysis and their resolutions and measured rms are summarized in Table\,\ref{tab:observations}.

\begin{table}
    \centering
    \caption{Observations used in this analysis}
    \begin{tabular}{c|c|c|c|c}
        \hline
        \hline
         Region & Line & Beam & RMS & Velocity\\
          & & (\arcsec) & (K) & Resolution  (km/s) \\
         \hline
         Ridge & \thirtCO(1-0) & 13 & 0.03  & 0.5 \\
         Ridge & CS(2-1) & 18 & 0.017 & 0.5 \\
         30~Dor & \thirtCO(2-1) & 13 & 0.009 & 0.25 \\
         N159 & \thirtCO(1-0) & 13 & 0.04 & 0.5 \\
         N113 & \thirtCO(1-0) & 13 & 0.0075 & 0.5 \\
    \end{tabular}
    \label{tab:observations}
\end{table}

\subsection{Molecular Ridge} \label{subsec:ridge}

\begin{figure*}
    \centering
    \includegraphics[width=0.45\textwidth]{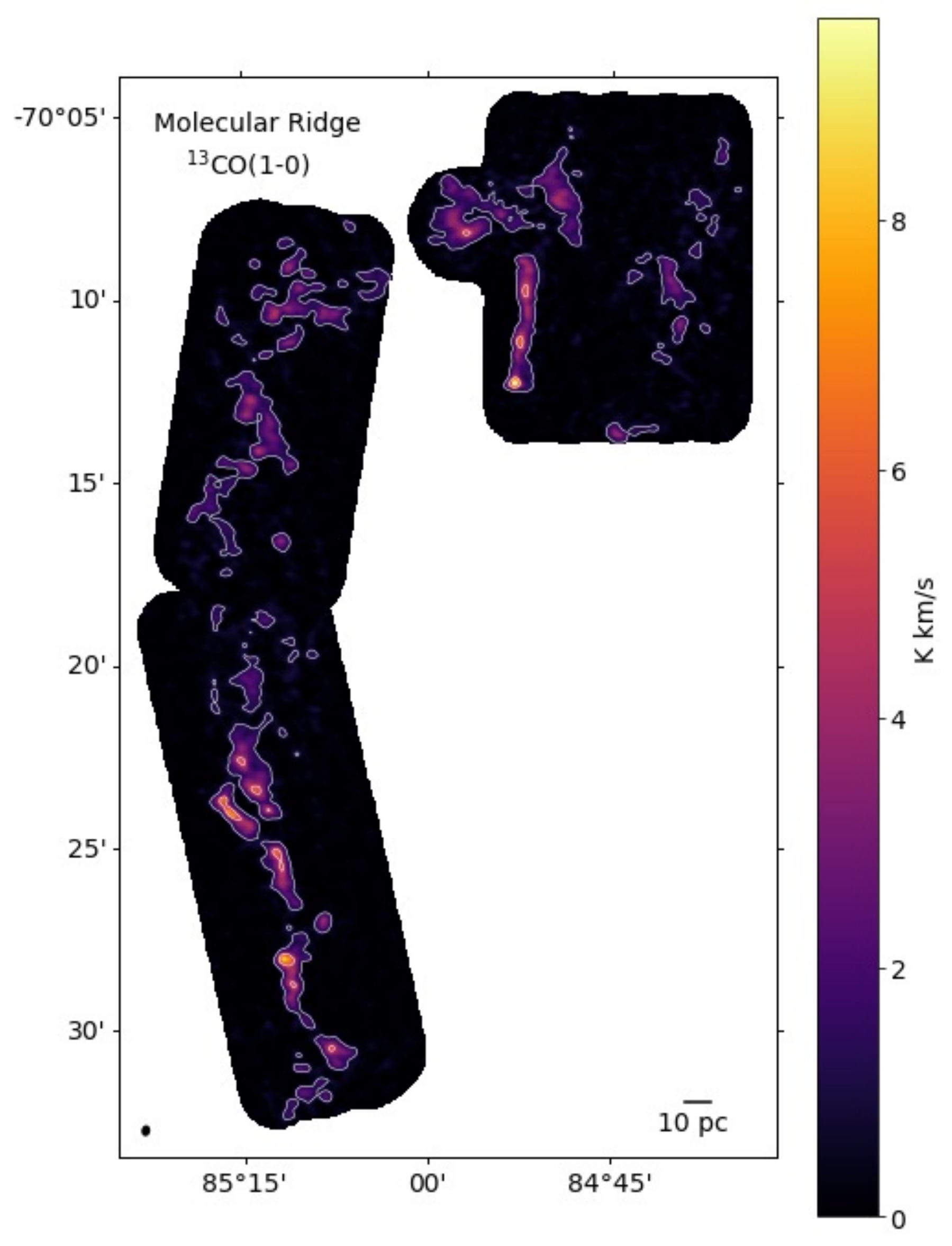}
    \includegraphics[width=0.47\textwidth]{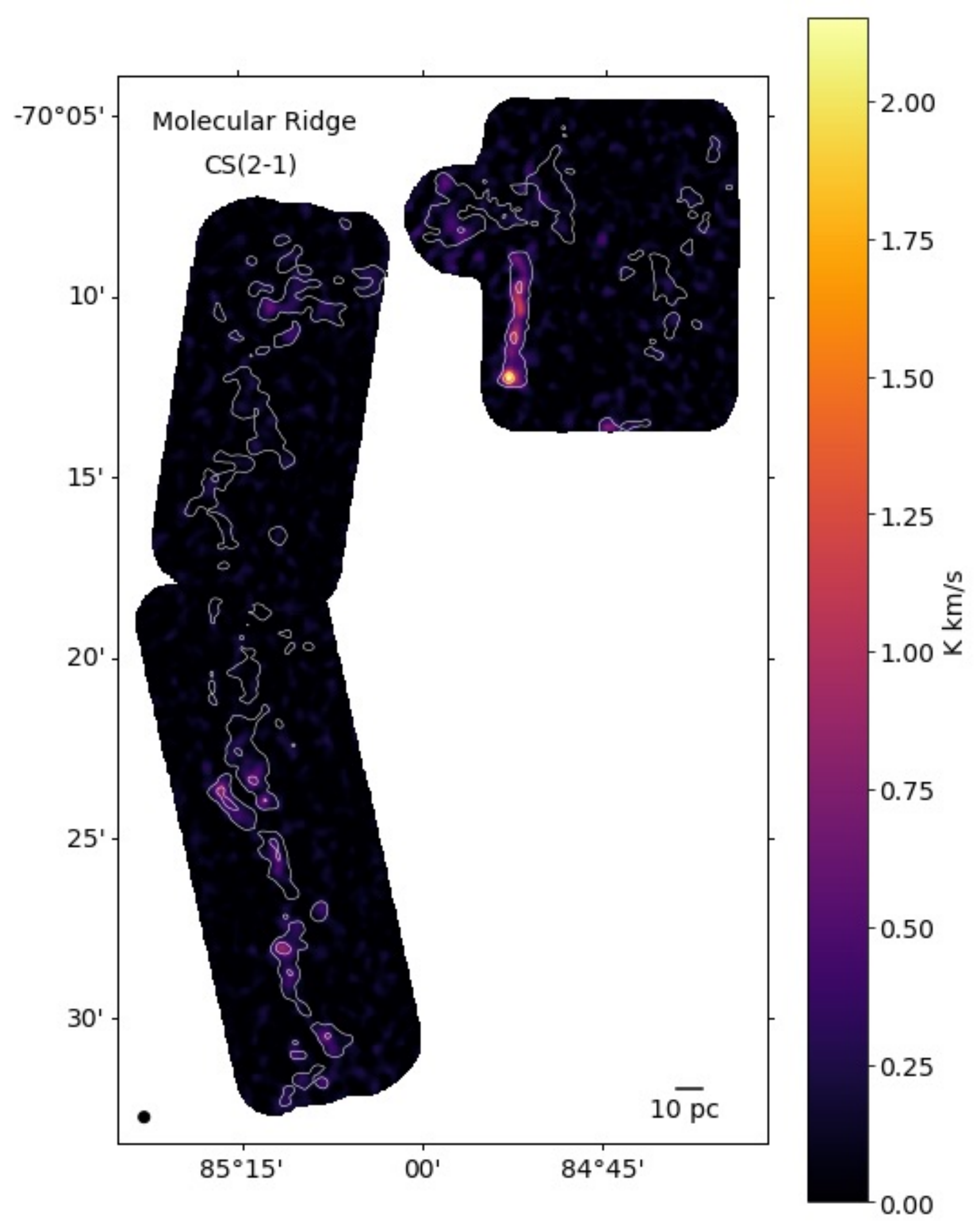}
    \caption{Left: Integrated intensity map of the Molecular Ridge in \thirtCO(1-0). Contours show 1, 5, and 9 K km/s.  Right: Integrated intensity map of the Molecular Ridge in CS(2-1), with the same contour lines from \thirtCO(1-0) overplotted. In both images, the beam is shown in the lower left corner.}
    \label{fig:ridge mom0}
\end{figure*}

The Molecular Ridge was observed by the Atacama Large Millimeter/submillimeter Array (ALMA) 7m Atacama Compact Array (ACA) in three maps for project 2017.1.00271.S. This data set was combined with ALMA 7m ACA projects 2012.1.00603.S and 2015.1.00196.S, which covered the $\sim2$ arcminute region around 5:39:50 -70:08:00 in the {northern} center of the maps. These were all also combined with ALMA total power data. The data reduction, calibration, and imaging are described in detail in \cite{Finn21}. The final maps have a resolution of 13\arcsec\ and a measured rms in line-free channels of 0.03 K. An integrated intensity map of the \thirtCO(1-0) and CS(2-1) emission is shown in Figure\,\ref{fig:ridge mom0}.

\subsection{30~Doradus} \label{subsec:30dor}

\begin{figure}
    \centering
    \includegraphics[width=0.47\textwidth]{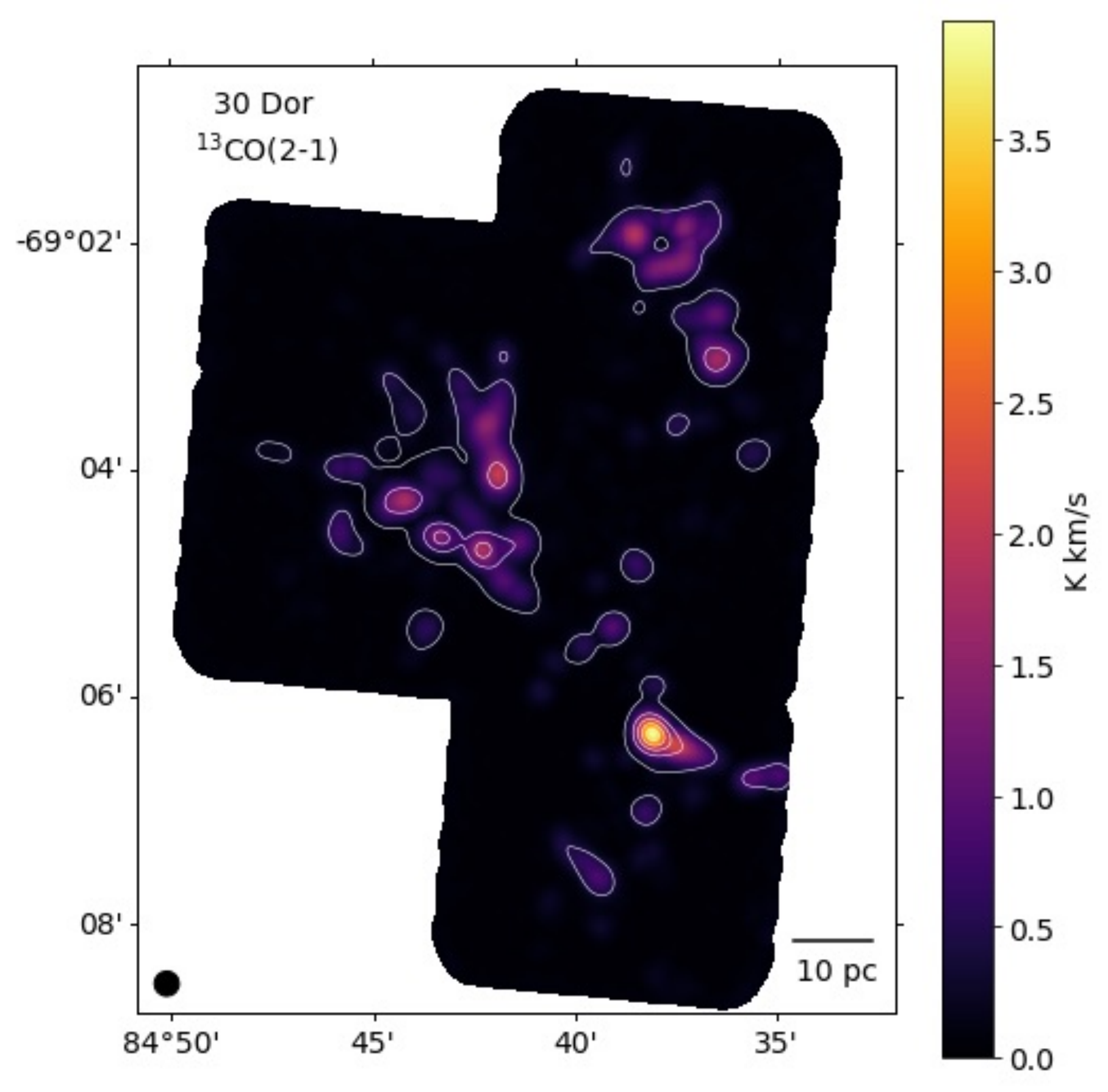}
    \caption{Integrated intensity map of 30~Dor in \thirtCO(2-1). Contours show 1, 5, 9, 13, and 17 K km/s. The beam is shown in the lower left corner.}
    \label{fig:30dor mom0}
\end{figure}

A mosaic of 30~Dor was observed as ALMA project 2019.1.00843.S, and includes 12m and 7m interferometric data as well as total power data.  Those data are presented  and analyzed at their native resolution of 1.75\arcsec\ in \cite{Wong22}. For this analysis, the resolution has been convolved to 13\arcsec\ so that it can be directly compared with our data for the Molecular Ridge. After this convolution, the data have a measured rms in line-free channels of 0.009 K. This rms is much lower than the other datasets, but we find that removing structures below the Ridge's noise level of 0.03 K from the analysis does not significantly change the results. An integrated intensity map of these data is shown in Figure\,\ref{fig:30dor mom0}.

This dataset is \thirtCO(2-1), rather than \thirtCO(1-0) like all the other regions. We use a \thirtCO(2-1)/\thirtCO(1-0) ratio of 0.84 based on the \twelveCO(2-1)/\twelveCO(1-0) ratio measured in \cite{Sorai01}.

\subsection{N159} \label{subsec:n159}

N159 is a massive star-forming region to the north of the Ridge but south of 30~Dor.  The brightest point in early single-dish surveys of CO in the LMC, this massive star formation region is less evolved than 30~Dor and has a significant remaining reservoir of molecular gas. N159 can be separated into eastern an western components (N159E and N159W), where N159E is thought to be more evolved than N159W \citep{Nayak18}. There is also another region to the south called N159S that is much more quiescent and is not included in this study. Due to the small number of structures identified at this work's resolution, we treat N159E and N159W as a single region and this does not significantly affect the results of our analysis.

\begin{figure}
    \centering
    \includegraphics[width=0.47\textwidth]{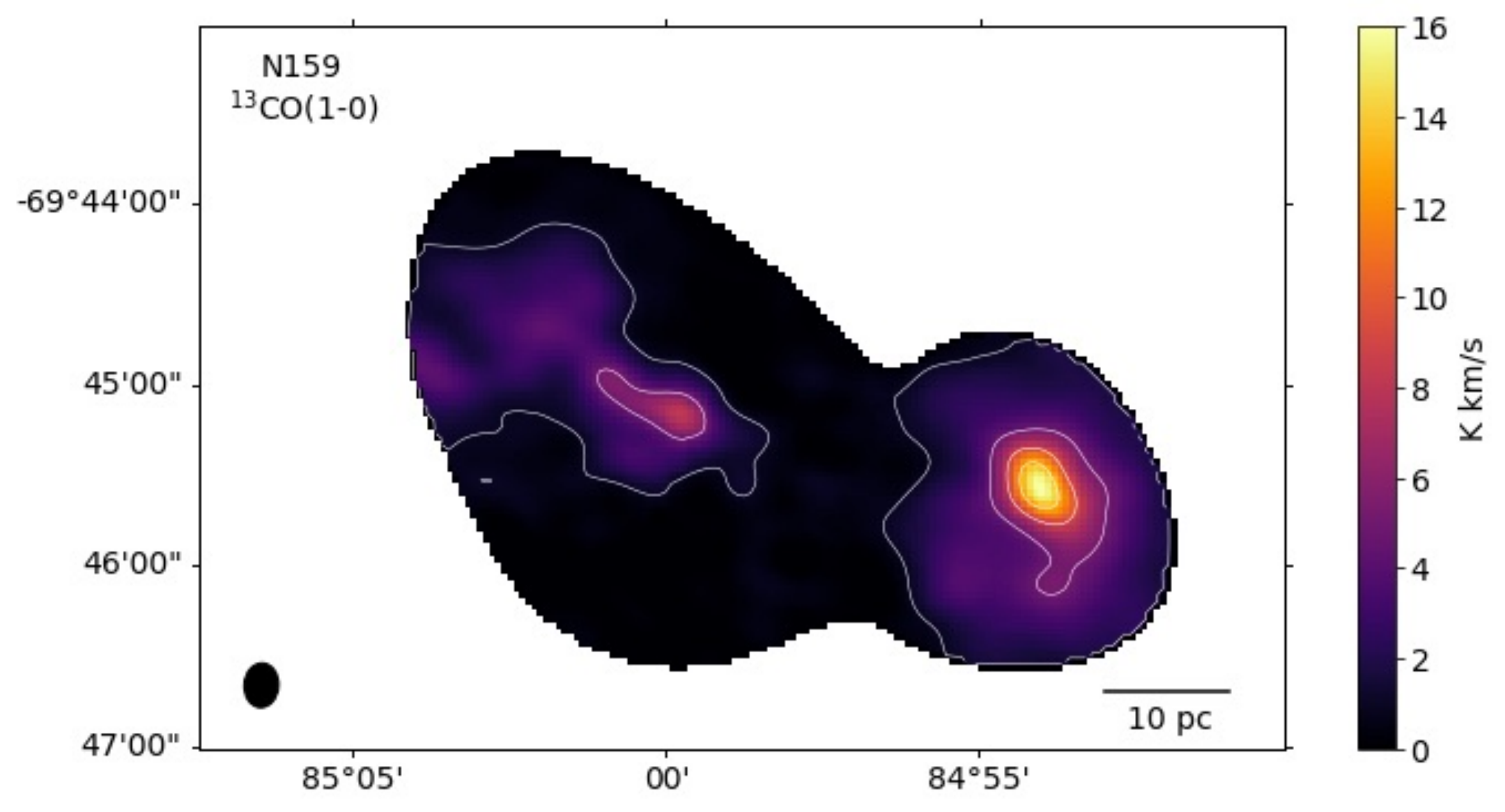}
    \caption{Integrated intensity map of N159 in \thirtCO(1-0). Contours show 1, 5, 9, and 13 K km/s. The beam is shown in the lower left corner.}
    \label{fig:n159 mom0}
\end{figure}

N159 was observed in \thirtCO(1-0) by ALMA project 2012.1.00554.S with 12m and 7m interferometric data, and was presented at the native resolution of 2.5\arcsec$\times$1.8\arcsec\ in \cite{Fukui2015}. These data do not include total power data, and so we expect a flux recovery around 66\% based on data in the Ridge. We take this correction into account in the mass estimates. As with the other comparison datasets, we convolved the data to 13\arcsec\ to match the Ridge data. After this convolution, the data have a measured rms in line-free channels of 0.04 K. An integrated intensity map of these data is shown in Figure\,\ref{fig:n159 mom0}.

\subsection{N113} \label{subsec:n113}

\begin{figure}
    \centering
    \includegraphics[width=0.47\textwidth]{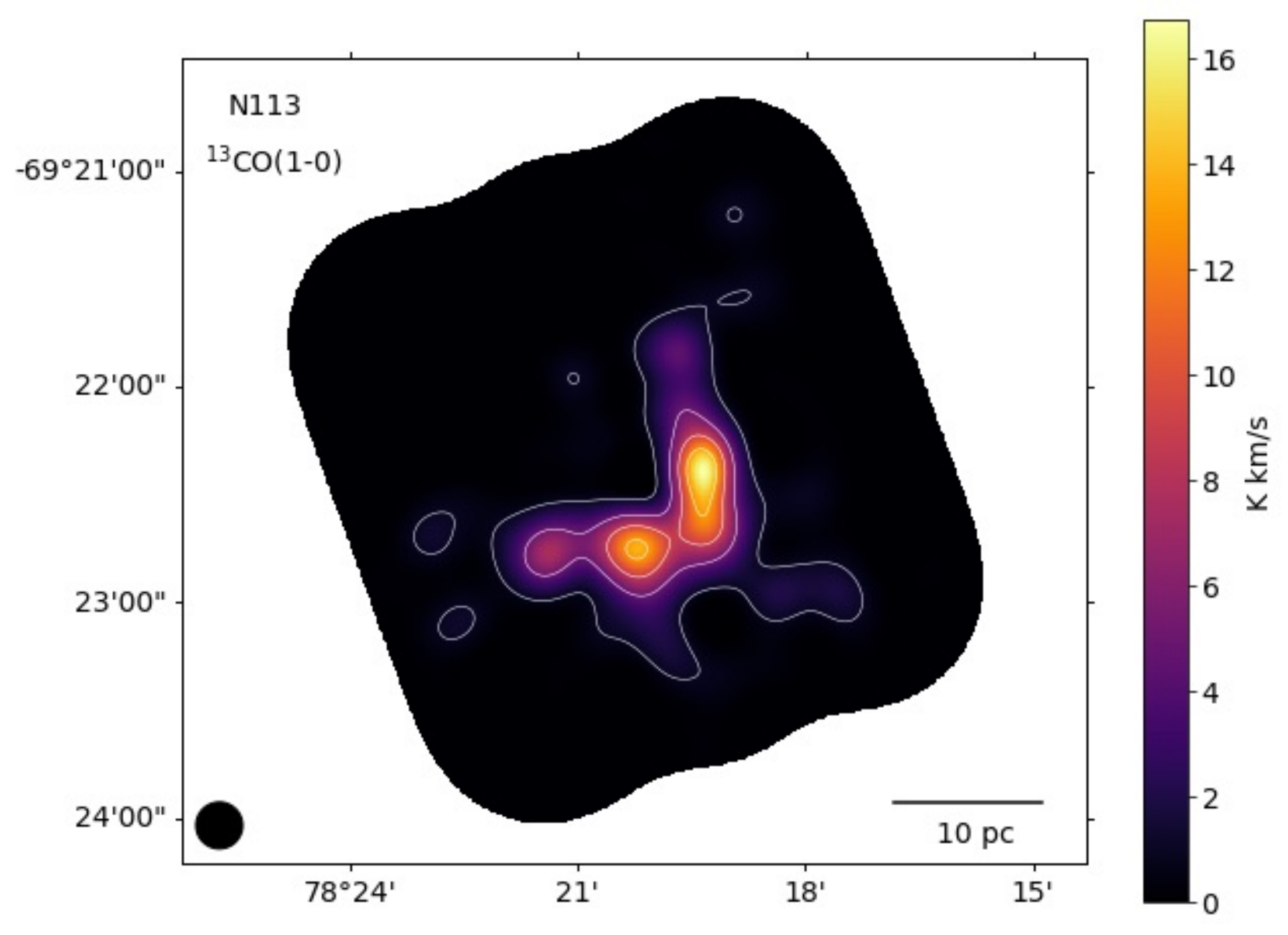}
    \caption{Integrated intensity map of N113 in \thirtCO(1-0). Contours show 1, 5, 9, and 13 K km/s. The beam is shown in the lower left corner.}
    \label{fig:n113 mom0}
\end{figure}

N113 is another active massive star-forming region in the LMC with several young, embedded massive YSOs. It is located in the central region of the LMC. N113 was observed in \thirtCO(1-0) by ALMA project 2015.1.01388.S with 12m interferometric data. 
These data do not include total power data, and so we expect a flux recovery around 66\% based on data in the Ridge. We take this correction into account in the mass estimates. A complete description of the data processing will be discussed in a separate publication at the observations' native resolution of $\sim2$\arcsec. For this work, the data were convolved to 13\arcsec\ so they could be directly compared to our Molecular Ridge data. After this convolution, the data have a measured rms in line-free channels of 0.0075 K. This rms is much lower than the other datasets, but we find that removing structures below the Ridge's noise level of 0.03 K from the analysis does not significantly change the results. An integrated intensity map of these data is shown in Figure\,\ref{fig:n113 mom0}.

\section{Structure Decomposition} \label{sec:dendrogram}

To decompose the emission into structures, we used two different methods: splitting the emission into a hierarchy of structures called a dendrogram, and splitting the emission into individual, non-overlapping clumps. Dendrograms better capture the complex and hierarchical nature of molecular clouds, but complicate analysis because the emission is multiply counted and the resulting structures often defy the physics commonly used to describe molecular clouds (for example, we report a single radius for molecular clouds, even though dendrogram structures are frequently non-spherical and have complex and elongated shapes). Clump finding algorithms offer a simpler approach to analysis, but are biased towards finding clump structures that are approximately round and the size of the beam, and they cannot capture the hierarchical structure of molecular clouds. 

In this study, we use both types of structure decomposition depending on the type of analysis being performed. We use dendrogram structures for the majority of the analysis since they better capture the hierarchical nature of clouds and can demonstrate how the physical properties behave at different size scales. We use clumps for counting-based analyses, such as histograms, since dendrograms multiply count emission. 

Due to its bias towards beam-sized structures, the clump segmentation method results in smaller ranges of masses, sizes, and linewidths being calculated in Section\,\ref{sec:properties}. The ranges of these parameters for the identified clumps align most closely with those for the leaves of the dendrogram.

\subsection{Dendrogram Segmentation} \label{subsec:dendro}

We decompose the emission in each map into structures using the package \texttt{astrodendro} \citep{Rosolowsky08} to create a dendrogram.  
This method of structure decomposition considers how different structures within the data merge as you go to lower contour levels to create a hierarchical categorization of the structures. We used input parameters of \texttt{min\_value}=3$\sigma$, \texttt{min\_delta}=2.5$\sigma$, and \texttt{min\_npix}=2 beams, meaning that the algorithm includes only pixels that are above $3\sigma$ and finds local maxima that are at least $2.5\sigma$ above the point of merging with another structure and bounded by an isosurface with at least as many voxels as two resolution beams. 

\begin{figure*}
    \centering
    \includegraphics[width=0.9\textwidth]{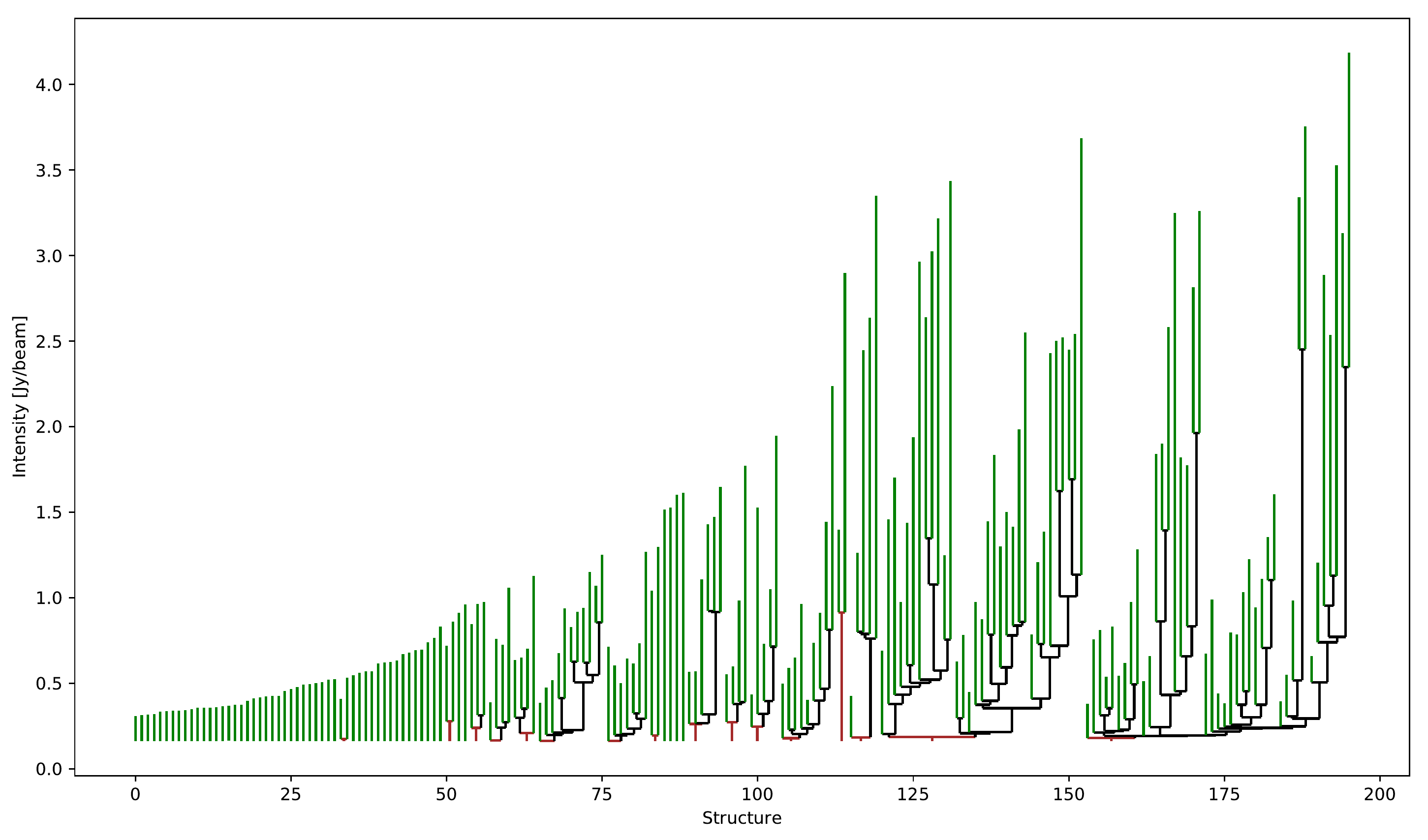}
    \caption{Dendrogram for the Ridge created by \texttt{astrodendro}. Each structure is represented by a vertical line, where the green lines are leaves, black lines are branches, and red lines are trunks. The y-axis indicates the peak of each structure and the intensity at which the structures merge with one another. This plot shows that the Ridge is composed of a few large trunks with a lot of substructure, but also several smaller and more isolated structures.}
    \label{fig:dendro}
\end{figure*}

The local maxima with no resolved substructure are categorized as ``leaves''. The algorithm then identifies the points at which these structures merge to define larger structures, categorized as ``branches'' and ``trunks'', where trunks are the largest structures that are not bounded by any other structures. Figure\,\ref{fig:dendro} shows the dendrogram for emission in the Ridge using the above parameters and the breakdown in dendrogram structures for each region is shown in Table\,\ref{tab:segmentation}.

\subsection{Clump Segmentation} \label{subsec:clump}

To decompose emission into clumps, we use the algorithm \texttt{quickclump} \citep{Sidorin17}. We used the input parameters \texttt{Nlevels}=1000, \texttt{Tcutoff}=4$\sigma$, \texttt{dTleaf}=4$\sigma$, and \texttt{Npixmin}=50. The resulting numbers of clumps for each region are shown in Table\,\ref{tab:segmentation}.

\begin{table*}
    \centering
    \caption{Results of dendrogram and clump segmentation, and fractal dimension, $D_2$, for each region}
    \begin{tabular}{c|c|c|c|c|c}
        \hline
        \hline
         Region & Trunks & Branches & Leaves & Clumps & $D_2$ \\
         \hline
         Ridge \thirtCO & 16 & 110 & 196 & 256 &  $1.50\pm0.02$\\
         Ridge CS & 4 & 6 & 34 & -- & -- \\
         30~Dor {\thirtCO} & 6 & 45 & 96 & 75 & $1.44\pm0.02$ \\
         N159 {\thirtCO} & 1 & 4 & 7 & 9 & --\\
         N113 {\thirtCO} & 1 & 6 & 10 & 10 &  $1.42\pm0.07$
    \end{tabular}
    \label{tab:segmentation}
\end{table*}

\subsection{Fractal Dimension} \label{subsec: fractal dimension}

\begin{figure}
    \centering
    \includegraphics[width=0.45\textwidth]{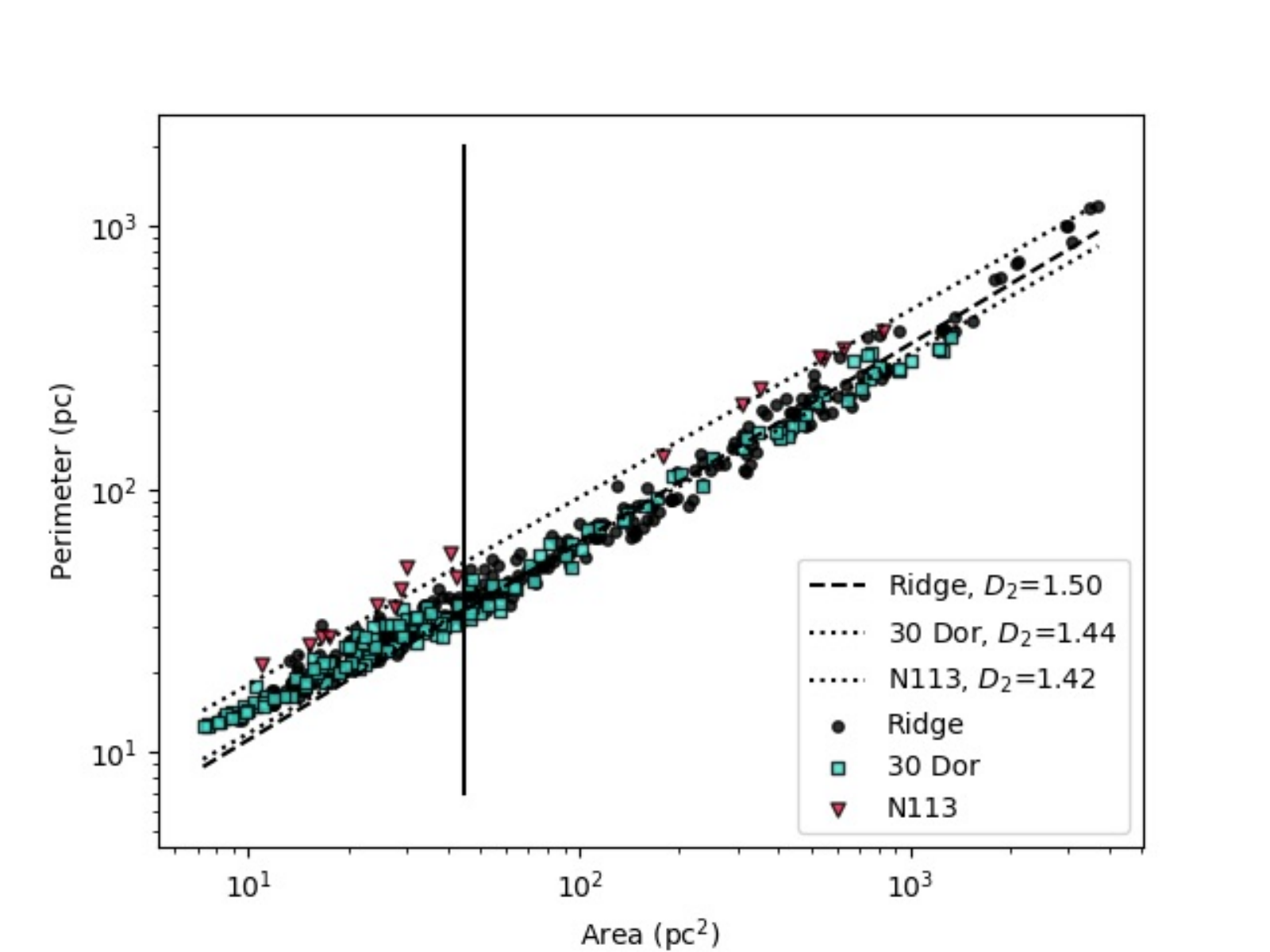}
    \caption{Perimeter plotted against area of the full contours for dendrogram structures identified in the Ridge, 30~Dor, and N113. We fit power laws to each region to measure the fractal dimension, $D_2$, with the area-perimeter relation, $P \propto A^{D_2/2}$. We fit only structures that are larger than the area of the beam, 45 pc, represented by the vertical line.  Each of the datasets are consistent with one another, suggesting that they have a similar hierarchical morphology. }
    \label{fig:fractal dimension}
\end{figure}

It is widely found that molecular gas is {encountered} in fractal structures \citep[i.e., ][]{ElmegreenFalgarone96}, and that this fractal nature is connected to supersonic turbulence in the interstellar medium \citep{Elmegreen01}.  We consider whether the emission in the Ridge has a {fractal morphology that is similar to the morphology in} 30~Dor and N113 by measuring the fractal dimension, $D_2$, of the structures identified by the dendrogram segmentation in each region. We do this with the area-perimeter relation, $P \propto A^{D_2/2}$ \citep{Falgarone91}. We measure the perimeter and the area of each structure based on the full contour defined by the \texttt{astrodendro} algorithm for the Ridge, 30~Dor, and N113, then fit a power law to each dataset using a non-linear least squares method \citep[\texttt{scipy.optimize.curve\_fit; }][]{scipy}. We do not include N159 because the emission on the edges is cut off by the observed map. We fit only the structures that have an area larger than the size of the beam, which is 45 pc$^2$. The resulting fits are shown in Figure\,\ref{fig:fractal dimension}. 

Each of the datasets are consistent with one another given the error, having fractal dimensions of $D_2 = 1.50\pm0.02$, $1.44\pm0.02$, and $1.42\pm0.07$ for the Ridge, 30~Dor, and N113, respectively. This suggests that the three regions have a similar hierarchical morphology and so likely have similar mechanisms by which turbulence regulates cloud structure.

These values are higher than the $D_2=1.36\pm0.02$ measured in galactic molecular clouds with \twelveCO\ by \cite{Falgarone91}, but are consistent with the range of 1.2-1.5 measured for HI emission in galactic clouds by \cite{Sanchez07}. These values are also consistent with similar measurements made using stellar structures in the LMC \citep{Miller22} and the Small Magellanic Cloud \citep[SMC; ][]{Sun18}, where both find $D_2=1.44\pm0.2$.

\section{Derived Properties} \label{sec:properties}

To compare the physical conditions of the molecular clouds in each region, we compute the mass, linewidth, radius, and virial parameter, \alphavir for each structure in each region.

We calculate the mass using an \XthCO factor to convert from integrated \thirtCO(1-0) intensity to H$_2$ column density. We adopt a value for \XthCO of 1.6$\times10^{21}$ cm$^{-2}$/(K km s$^{-1}$) based on \cite{Finn21} measurements of non-LTE model-fitted column density in the Ridge. \cite{Finn21} found that the non-LTE fitted column density of clumps was tightly correlated with the \thirtCO(1-0) integrated emission, even more so than with the \twelveCO(1-0) integrated emission. This value of \XthCO is also consistent with using a typical galactic \XCO factor of 2$\times10^{20}$ cm$^{-2}$/(K km s$^{-1}$ \citep{Bolatto13} and a \thirtCO(1-0)/\twelveCO(1-0) integrated intensity ratio of 0.12 \citep{Finn21}. The resulting column densities for 30~Dor also match the ranges of those calculated using LTE assumptions in \cite{Wong22}. 

We adopt an error of 10\% on the calculated masses based on an assumed 10\% flux calibration error \citep{Fomalont14}. The error associated with the measured $\sigma_\text{rms}$ of the image is negligible compared to the flux calibration error. Adopting an \XthCO factor also comes with a large systematic error. \cite{Finn21} cite an error of 50\% based on systematic uncertainties in the H$_2$/\thirtCO\ abundance ratio, which is added in quadrature to the 10\% flux error. The structures in 30~Dor have an additional systematic error since we only have observations of \thirtCO(2-1) and so we use a \thirtCO(2-1)/\thirtCO(1-0) ratio of 0.84$\pm$0.3 \citep{Sorai01}. 

To calculate the linewidth of each structure, we fit a Gaussian to the intensity-weighted mean line profile to determine $\sigma_v$ {(not the full-width at half-maximum, FWHM)}. This linewidth is then deconvolved with the velocity resolution of the data sets (either 0.5 or 0.25 km s$^{-1}$, see Table\,\ref{tab:observations}). To find the error in the measured linewidths, we create 100 noise maps that have been convolved to the same beam size and have the same $\sigma_\text{rms}$ as the emission maps and add these to the data then recompute the linewidth. The error is taken to be the standard deviation in the measured linewidths with the added noise. 

We performed the above method of error calculation for N159, N113, and 30~Dor, but this was not computationally feasible for the Ridge due to the large size of the data cube and number of clumps. Instead, we found that the fractional error in $\sigma_v$ ($\sigma_{\sigma_v}/\sigma_v$) from this method in N159, N113, and 30~Dor {is} closely correlated with the peak {brightness temperature} ($T_\text{max}$) of the structure. We fit this correlation and extrapolated it to the Ridge data set to determine approximate errors for each structure using the following fitted equation:

\begin{equation}
    \log{\left(\frac{\sigma_{\sigma_v}}{\sigma_v}\right)} = -0.69\log{(T_\text{max}/K)} - 2.23.
\end{equation}

To calculate the sizes, we fit an ellipse to the half-power contour of the structure. To get a single radius value, we used the geometric mean of the major and minor axes of this fitted ellipse, which is then deconvolved with the beam size for the data set. This value is taken to be a HWHM of the structure, from which we approximated $\sigma_R=$HWHM$\times2/2.35$. We then multiplied $\sigma_R$ by a factor of 1.91 \citep{Solomon87} to get our quoted ``effective radius'', $R$. The error was determined with the same method used for the error in the linewidth. In the case of the radius, the fractional errors from N159, N113, and 30~Dor are also closely correlated with the peak {brightness temperature} ($T_\text{max}$) of the structure, and this fitted correlation was extrapolated to the Ridge to approximate errors with the equation 

\begin{equation}
    \log{\left(\frac{\sigma_{R}}{R}\right)} = -1.09\log{(T_\text{max}/K)} - 1.60.
\end{equation}

Deconvolving the radius with the beam and the linewidth with the velocity resolution resulted in some data points being dropped from the analysis because the half-power contour is smaller than the beam or the fitted linewidth is smaller than the velocity resolution. This left us with 204 structures in the Ridge, 51 structures in 30~Dor, and 9 structures each in N159 and N113.

{For N159 and N113, we adopt an additional correction factor of 1/0.66 in the measured masses to account for the lack of total power data, which we expect to result in a flux recovery of 66\% based on data from the Ridge that includes total power. We find that leaving out the total power data in the Ridge does not significantly affect the measured radii or linewidths of the structures. }

The three parameters above, mass ($M$), radius ($R$), and linewidth, $\sigma_v$, are used to calculate the virial parameter, \alphavir, as a measurement of the balance between gravity and outward pressure, {calculated for a spherical cloud. Most molecular cloud structures that we observe are not spherical, which could influence the value of \alphavir by a factor of order unity. Values of \alphavir greater than one indicate that the cloud is dominated by kinetic energy, which could mean that the cloud is not bound and will disperse, or that it is constrained by an external pressure to keep it bound. The kinetic energy could also be dominated by potential energy because the cloud is in free fall collapse. Values of \alphavir less than one indicate that the cloud is dominated by potential energy, and so is likely to begin collapse.  } We use the equation

\begin{equation}
    \alpha_\text{vir} = \frac{5 \sigma_v^2 R}{GM}.
\end{equation}

The parameters derived in this section are used in the following sections for the remainder of the analysis. {Throughout the analysis, we use the properties derived from \thirtCO\ emission rather than CS unless specified otherwise. }

\section{Size-Linewidth Relations} \label{sec:SL plots}

We plot the {\thirtCO\ } linewidths of all the structures against their effective radii. The relation between the two is expected to follow a power law \citep{Larson81, Solomon87} of the form

\begin{equation}
    \sigma_v = a_0 R^{a_1}.
\end{equation}

\cite{Solomon87} measured sizes and linewidths for molecular clouds in the Galactic disk using a size parameter, $S$, instead of the effective radius, $R$, that we use in this study. The size parameter is the geometric mean of the spatial dispersions, $\sigma_l$ and $\sigma_b$, of each cloud, and so is comparable to the $\sigma_R$ that we measured and then converted to effective radius with the equation $R = 1.91\sigma_R$ (see discussion in \S\ref{sec:properties}). \cite{Solomon87} fit values of $a_0 = 1.0\pm0.1$ and $a_1 = 0.5\pm0.05$, so converting their size parameter to an effective radius would result in an intercept for the power law of $a_0=0.72\pm0.07$. 

\begin{figure}
    \centering
    \includegraphics[width=0.45\textwidth]{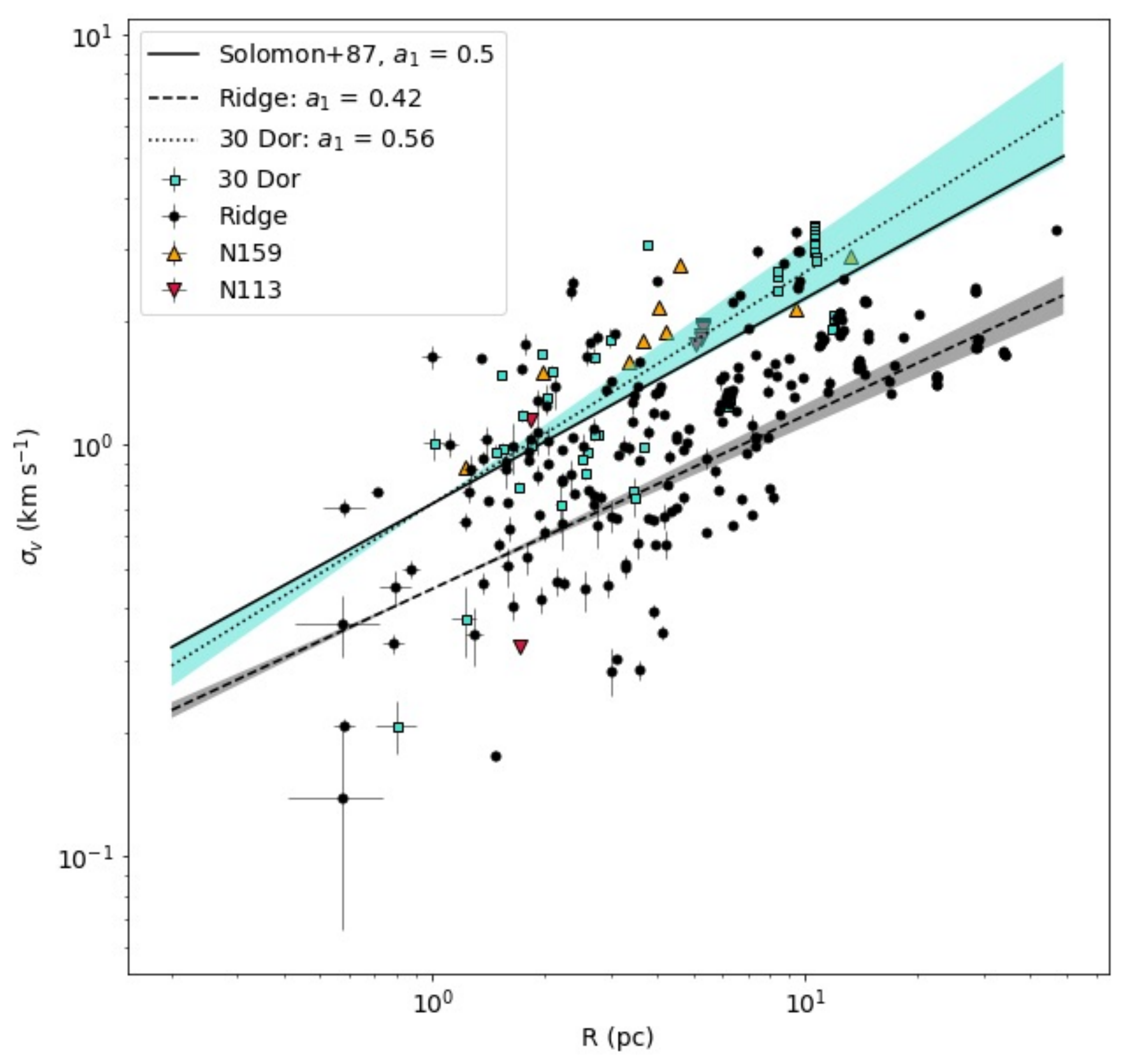}
    \caption{Linewidths plotted against sizes of structures in the four different regions: the Ridge (black circles), 30~Dor (blue squares), N159 (orange upward-facing triangles), and N113 (red downward-facing triangles). A power law has been fit to the structures in the Ridge (dashed line) and 30~Dor (dotted line) with the error in the fit shown as shaded regions, and the value of the fitted power law slope, $a_1$, is written in the legend for each region. We do not fit power laws for N159 and N113 since their small numbers of data points cannot constrain the parameters well. The relation fit by \cite{Solomon87} for Milky Way clouds is shown as a solid line, where $a_0 = 0.72$ and $a_1 = 0.5$. {This indicates that 30~Dor follows a steeper power law than the Ridge, so it may have more kinetic energy at larger size scales than the Ridge, although the two values of $a_1$ are within 3$\sigma$ of both each other and the fit by \cite{Solomon87}.} }
    \label{fig:SL slope}
\end{figure}

In Figure\,\ref{fig:SL slope}, we plot the radius and velocity dispersion of the structures and fit both the intercept and slope, $a_0$ and $a_1$, for the Ridge and 30~Dor, using an orthogonal distance regression to take into account the error in both axes \citep[\texttt{scipy.odr; }][]{scipy}. We do not fit these values for N159 and N113 since there are so few data points and the fits are poorly constrained. 

For the Ridge, we fit values of $a_0 = 0.41\pm0.03$ and $a_1 = 0.44\pm0.03$, while for 30~Dor we fit values of $a_0 = 0.72\pm0.11$ and $a_1 = 0.56\pm0.07$. This indicates that 30~Dor follows a  steeper power law than the Ridge, although the two values of $a_1$ are within 3$\sigma$ of each other. This means that 30~Dor may have more kinetic energy at larger size scales than the Ridge. The power law slope we fit for 30~Dor structures ($a_1 = 0.56\pm0.07$) is comparable to those fit by \cite{Nayak16}, \cite{Wong19}, \cite{Indebetouw20}, and \cite{Wong22} for dendrogram structures identified in \twelveCO(2-1) and \thirtCO(2-1) emission at resolutions ranging from 0.1 to 0.85 pc. 



\begin{figure}
    \centering
    \includegraphics[width=0.45\textwidth]{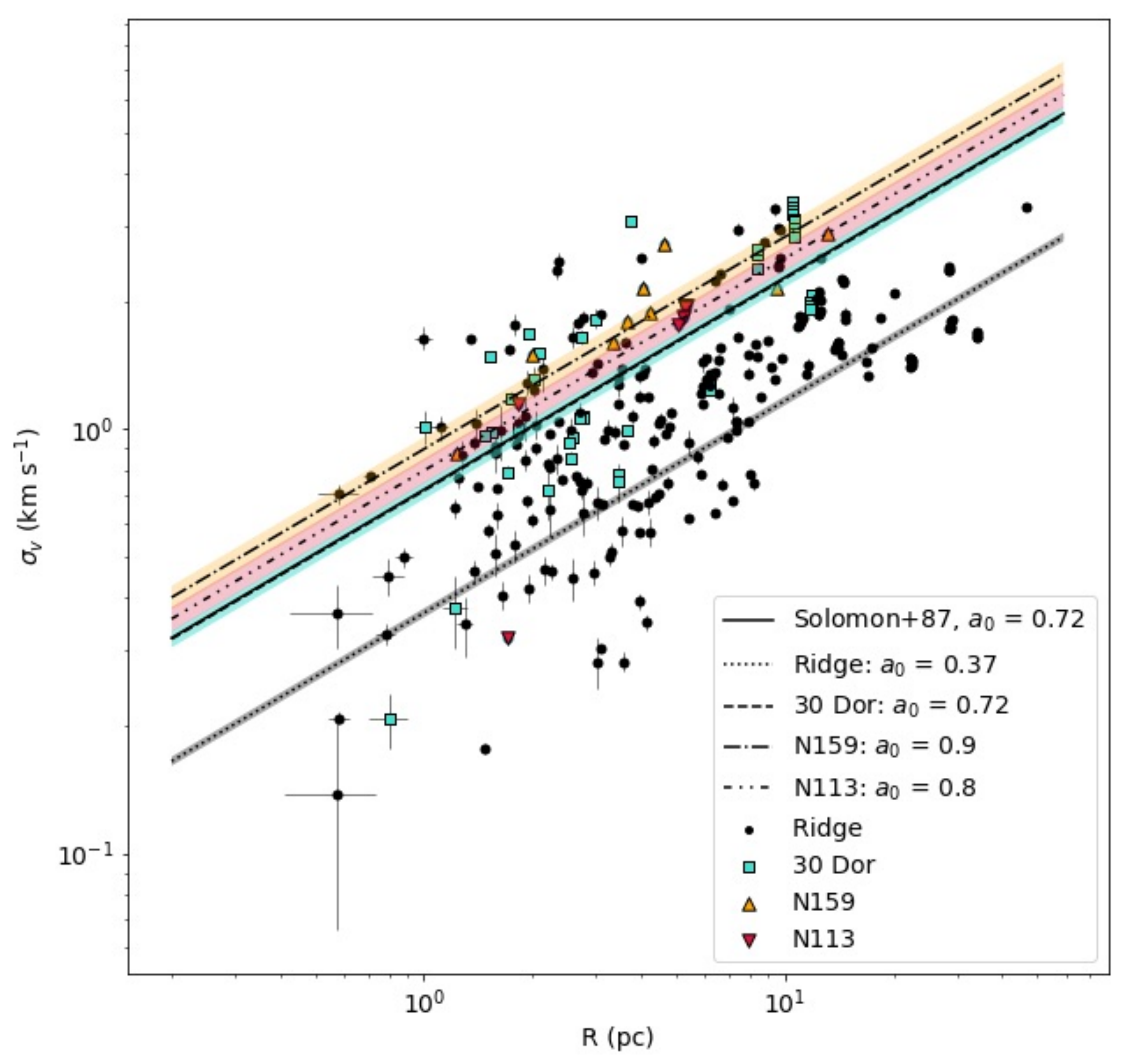}
    \caption{Linewidths plotted against sizes of structures in the four different regions: the Ridge (black circles), 30~Dor (blue squares), N159 (orange upward-facing triangles), and N113 (red downward-facing triangles). A power law with a fixed slope of $a_1 = 0.5$ has been fit to the structures in each region and the error on each fit is shown as shading colored according to the region, and the value of the fitted power law intercept, $a_0$, is written in the legend for each region. The relation fit by \cite{Solomon87} for Milky Way clouds is shown as a solid line, where $a_0 = 0.72$. The Ridge has a significantly lower fitted intercept than the other three regions, indicating that it has much less kinetic energy for a given size scale.}
    \label{fig:SL intercept}
\end{figure}

In Figure\,\ref{fig:SL intercept}, we {plot again} the size and linewidth of the structures, but now hold the slope of the power law fixed at $a_1 = 0.5$ \citep{Solomon87} and fit only the intercept, $a_0$. In this case, we fit a value of $a_0 = 0.35\pm0.01$ for the Ridge, $a_0 = 0.72\pm0.03$ for 30~Dor, $a_0 = 0.90\pm0.06$ for N159, and $a_0 = 0.80\pm0.05$ for N113. Changing the value of the fixed slope changes the fitted values of the intercept slightly, but does not change the relative differences between data sets. The results above are also unchanged if we use clump-segmented structures instead of dendrogram structures.

The Ridge's significantly lower intercept from the other three data sets indicates that it has less kinetic energy at a given size scale than the massive-star-forming regions. This result suggests that the lack of massive star formation in the Ridge cannot be caused by excess kinetic energy suppressing star formation. This is quite different from the situation in the Galactic Center, where star formation is also suppressed but the molecular clouds have much higher kinetic energies \citep[{$a_0 = 3.0\pm0.7$ after converting the size parameter to an effective radius}; ][]{Oka01} than clouds in the disk of the Galaxy \citep[$a_0 = 0.72\pm0.07$; ][]{Solomon87} . This indicates that while both regions are examples of deviations from star formation scaling laws, the physical drivers of that suppression are different. 

\subsection{Size-Linewidth with CS} \label{subsec: SL with CS}

\begin{figure}
    \centering
    \includegraphics[width=0.45\textwidth]{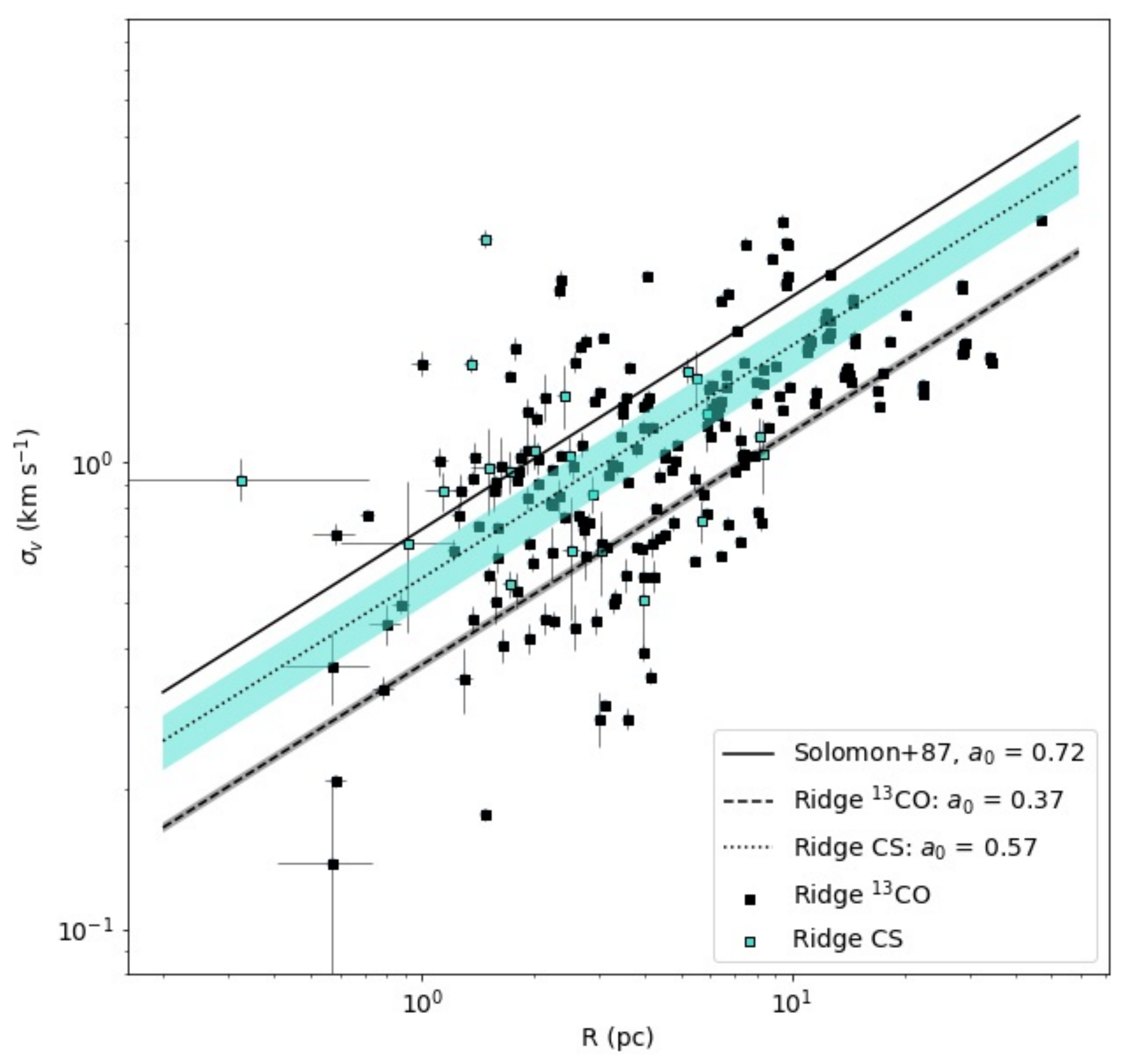}
    \caption{Linewidths plotted against the sizes of structures in the Ridge identified by \thirtCO(1-0) (black) and CS(2-1) (blue). We fit a power law with a fixed slope of $a_1 = 0.5$ to each, which shows that the CS(2-1) structures have a significantly higher fitted intercept than the \thirtCO(1-0) structures. This indicates that the dense gas traced by CS(2-1) has higher kinetic energy at a given size scale.}
    \label{fig:SL CS}
\end{figure}

In Figure\,\ref{fig:SL CS} we plot the size and linewidth of the structures in the Ridge in \thirtCO(1-0) and CS(2-1) and compare the fitted power law intercept with a fixed slope of $a_1 = 0.5$. The structures from CS(2-1) have a fitted intercept of $a_0 = 0.57\pm0.08$, higher than the \thirtCO(1-0) structure in the Ridge but within $3\sigma$ of the $a_0 = 0.35\pm0.01$ fitted above. This demonstrates that the dense gas structures traced by CS(2-1) in the Ridge have a higher kinetic energy than the more diffuse gas traced by \thirtCO(1-0). 

It is expected that measuring the sizes and linewidths in a dense gas tracer as opposed to \thirtCO\ would result in a size-linewidth relation with a higher intercept \citep{Goodman98}. However, it is not clear from this plot alone what is causing the higher energy levels in the case of the Ridge presented here. It could be because the gas traced by CS(2-1) is found at the dense centers of clumps throughout the Ridge, where the higher kinetic energy is balanced by {a} higher gravitational potential. Or it could be that the areas of the Ridge that have CS(2-1) detections are sites where there is more star formation and gravitational collapse occurring, and so those areas have higher kinetic energies than the rest of the Ridge. \cite{Finn21} see some correlation between the presence of YSOs and the CS(2-1)/\twelveCO(1-0) ratio in the Ridge, suggesting some support for the latter scenario, or that a combination of these two effects are at play.

\section{Virialization} \label{sec: virial plot}

\begin{figure}
    \centering
    \includegraphics[width=0.47\textwidth]{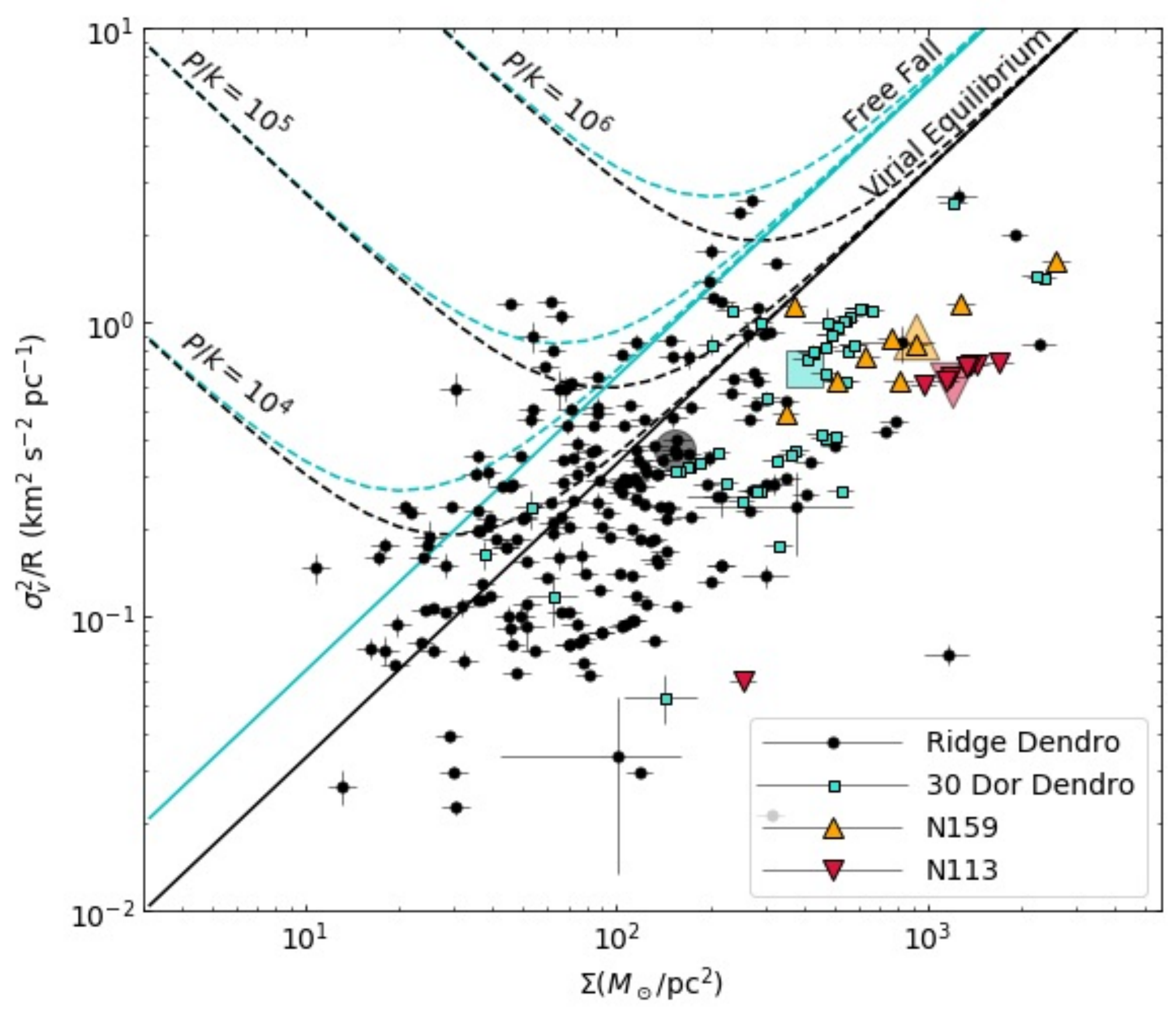}
    \caption{Surface density plotted against the velocity metric, $\sigma_v^2/R$, for the Ridge, 30~Dor, N159, and N113. The black line shows virial equilibrium, the blue line shows how equilibrium shifts when clouds are in free fall, and the dashed lines show those same quantities but when an external pressure is applied \citep{Field11}. The large, semi-transparent symbols corresponding to each region indicate the average surface density and velocity metric for each region. The systematic uncertainties for all data points are shown in the legend. The structures in the massive-star-forming regions appear to fall below the line of virial equilibrium more than the structures in the Ridge.}
    \label{fig:virial}
\end{figure}

In Figure\,\ref{fig:virial}, we plot $\sigma_v^2/R$ against the surface density of each structure. The positions of the structures on this plot indicate the balance between their gravitational potential energy and kinetic energy due to turbulence and temperature. If those two are in virial equilibrium, the structures should fall along the virial line, shown in black in Figure\,\ref{fig:virial}. 

Due to the large systematic uncertainties in the measurements of the structures' masses, the absolute positions relative to virial equilibrium on the plot are ambiguous. Instead, we focus this analysis on the relative positions of the data points from different regions. {The Ridge structures tend to fall above virial equilibrium more often than  structures in the massive-star forming regions 30~Dor, N159, and N113 - 28\% of the Ridge structures fall above the virial equilibrium line, where for 30~Dor that fraction is only 3\% and none of the structures in N159 or N113 fall above the virial equilibrium line. }

\begin{figure}
    \centering
    \includegraphics[width=0.45\textwidth]{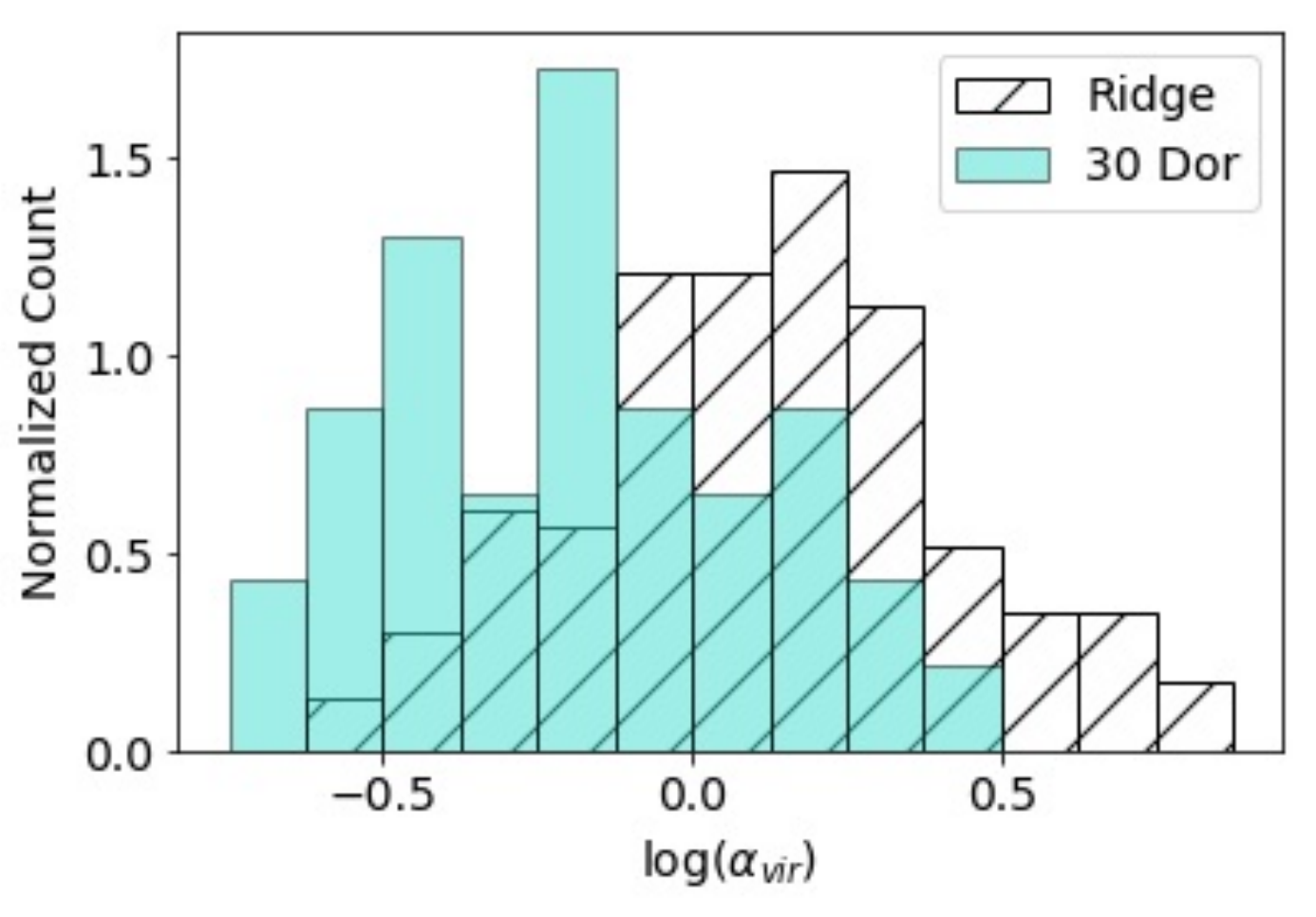}
    \caption{Distribution of \alphavir values in the Ridge and 30~Dor, showing that the Ridge tends towards higher values of \alphavir than 30~Dor. A K-S test indicates that the two datasets are not drawn from the same distribution with a p-value of $\ll0.01$. }
    \label{fig:alphavir hist}
\end{figure}

To more quantitatively compare the balance of kinetic and gravitational energy in the Ridge and 30~Dor, Figure\,\ref{fig:alphavir hist} shows the distribution of \alphavir values in the two regions. Because dendrograms multiply count emission, we use structures from \texttt{quickclump} to create all histograms in this analysis. Figure\,\ref{fig:alphavir hist} clearly shows a difference in \alphavir values between the two regions, with the Ridge structures tending towards higher \alphavir. A Kolmogorov-Smirnoff (K-S) test indicates that the two data sets are likely not drawn from the same distribution with a p-value of $\ll0.01$. 

\begin{figure}
    \centering
    \includegraphics[width=0.45\textwidth]{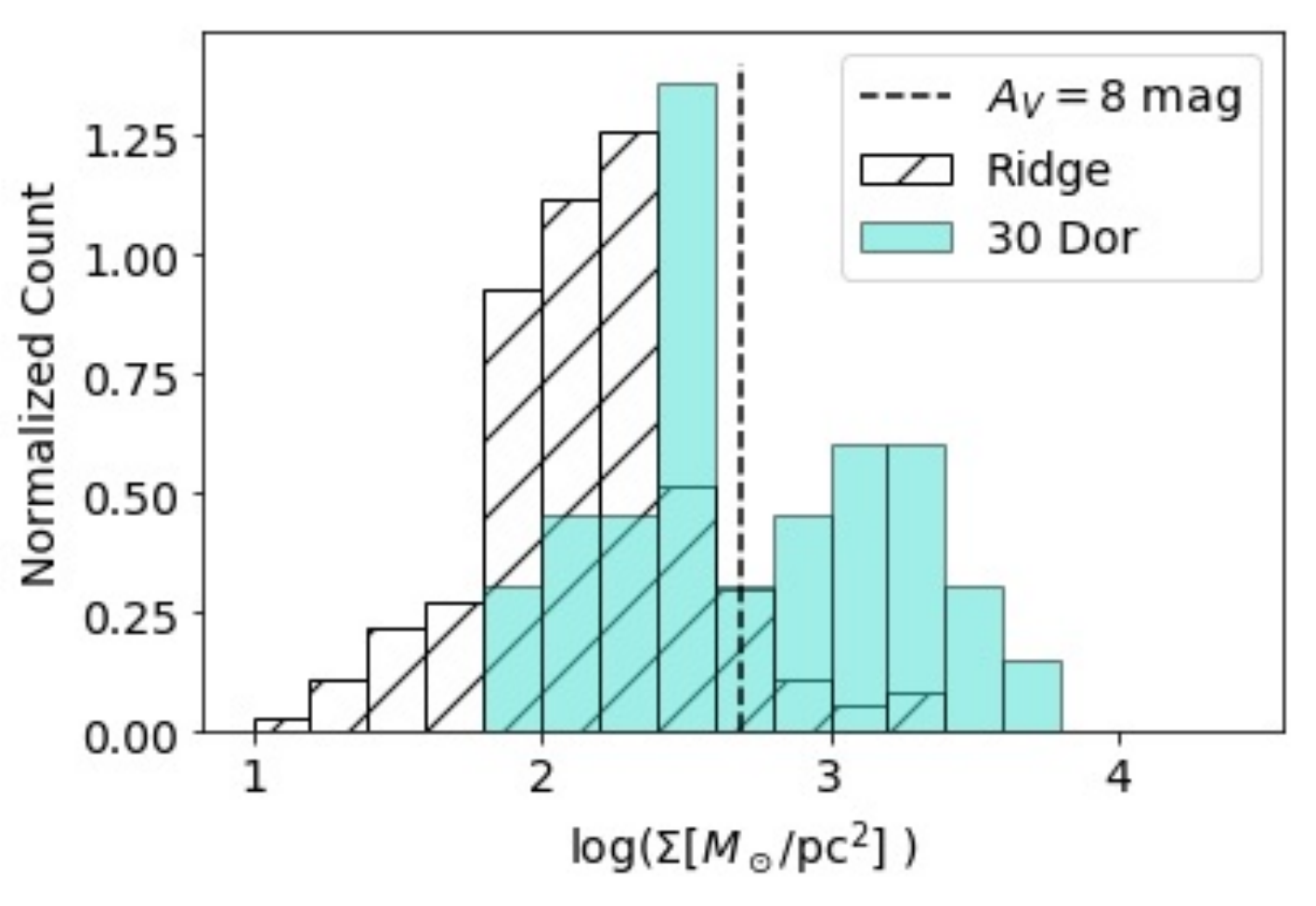}
    \caption{Distribution of surface densities in the Ridge and 30~Dor. The Ridge has significantly lower surface densities than 30~Dor, which results in structures in the Ridge having higher \alphavir values despite their lower kinetic energies. Also shown is the $A_V>8$ {mag} threshold for massive star formation from \cite{Lada10}, converted to $\Sigma_\text{gas} > 490$ M$_\odot$ pc$^{-2}$. This indicates that the Ridge is likely not forming massive stars because it lacks dense gas. }
    \label{fig:Sigma hist}
\end{figure}

We infer from Figure\,\ref{fig:SL intercept} that structures in the Ridge have lower kinetic energies than structures in 30~Dor. That means that for the Ridge to have higher \alphavir values than 30~Dor, it must have much lower surface densities. In Figure\,\ref{fig:Sigma hist}, we show the distributions of surface densities in the Ridge and 30~Dor, and we do indeed see that those in the Ridge are significantly lower than those in 30~Dor, with a K-S test p-value of $\ll0.01$. 
This suggests that the Ridge may be forming fewer massive stars compared to other regions because the Ridge does not have as much dense gas.

Given the much lower star formation activity in the Ridge and applying the Schmidt-Kennicutt law \citep{Kennicutt98}, it is expected that the surface density in the Ridge be lower than in 30~Dor. However, it is still higher than what would be predicted with the Schmidt-Kennicutt law based on the star formation rate measured in the Ridge by \cite{Indebetouw08}, $4\times10^{-3}$ M$_\odot$ yr$^{-1}$. From that rate, we would expect a surface density of $\sim47$ M$_\odot$ pc$^{-2}$, but here we measure a mass-weighted average surface density of $\sim100$ M$_\odot$ pc$^{-2}$ in the Ridge. This demonstrates that while the Ridge is lower in surface density than 30~Dor, there is still a discrepancy with the Schmidt-Kennicutt law in the Ridge.

We compare these measured surface densities to the often-cited threshold for massive star formation of $A_V > 8$ {mag} measured by \cite{Lada10}. \cite{Finn2019} calculate that based on measurements of $\frac{A_V}{N_{H}}$ in the LMC \citep{Dobashi08}, this threshold would correlate to $\Sigma_\text{gas} > 490$ M$_\odot$ pc$^{-2}$. Plotting this threshold on Figure\,\ref{fig:Sigma hist} indicates that the majority of structures in the Ridge fall below this threshold, while 30~Dor structures are distributed around this threshold and have a mass-weighted average of 500 M$_\odot$ pc$^{-2}$. There were too few structures in N159 and N113 at this resolution to include them in a histogram, but their mass-weighted average surface densities were 1100 and 1260  M$_\odot$ pc$^{-2}$, respectively. This is all consistent with the Ridge having little massive star formation because it falls below a density threshold compared to the other massive star forming regions.

{It is expected that highly irradiated clouds, such as those in 30~Dor close to R136, will have a higher dark-gas fraction due to photodissociation and so a higher \XCO factor \citep{Chevance20}. This effect would also cause us to see less of the diffuse envelope of the cloud, meaning that both the measured radii and linewidths may be underestimated. O'Neill et al. (submitted) make a thorough study of the effect of CO-dark gas on the measured properties of clouds. Their analysis suggests that in extreme cases of compactness for clouds in 30~Dor, the discrepancy in Figure\,\ref{fig:virial} could be explained by a significantly higher dark gas fraction than in the Ridge. However, the effect of CO-dark gas on the size-linewidth relation, such as is shown in Figure\,\ref{fig:SL intercept}, would be to move up and to the right, mostly parallel to the relation. This suggests that it is quite unlikely that the difference in fitted $a_0$ intercept values is due to differences in the amount of CO-dark gas in the regions. Differences in the abundance ratio of \thirtCO\ would cause similar effects, but the abundances have not been measured well enough to make any strong comment on this effect.} 

{The use of \thirtCO(2-1) in 30~Dor and \thirtCO(1-0) in the Ridge and the other regions likely also affects their relative distributions of \alphavir, even after applying a correction factor of \thirtCO(2-1)/\thirtCO(1-0) = 0.84 to the 30~Dor masses. Since we expect \thirtCO(2-1) to be fainter and not as well detected as \thirtCO(1-0), this probably has a similar effect to that of CO-dark gas, although a thorough study taking into account excitation would be necessary to draw any firm conclusions.} 


\section{Variation within the Ridge} \label{sec: ridge variation}

\begin{figure}
    \centering
    \includegraphics[width=0.45\textwidth]{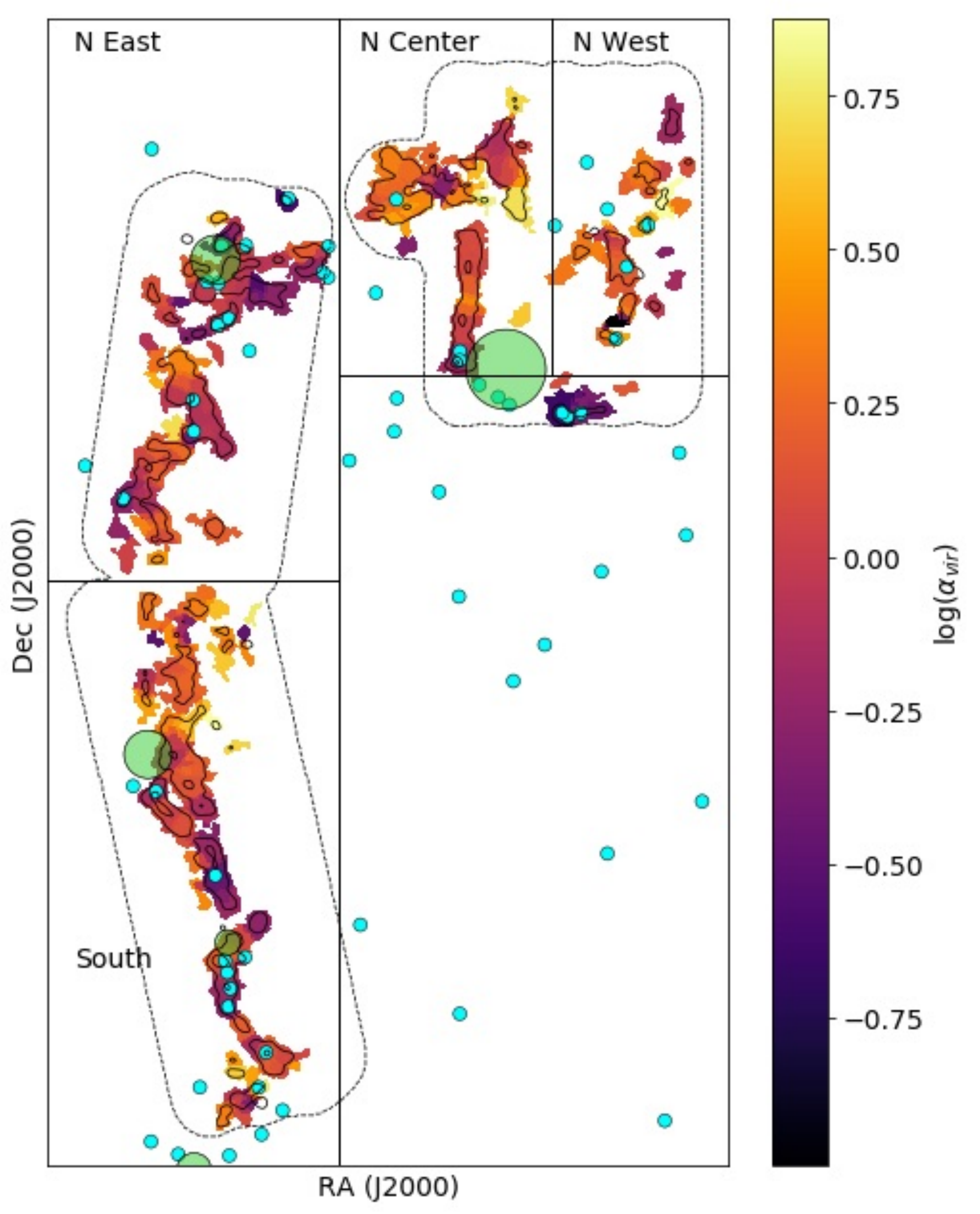}
    \caption{Map of the Molecular Ridge colored by each structure's \alphavir parameter. For this map we used structures found with \texttt{quickclump} to minimize structure overlap. Where there is overlap along the line of sight, we show the average \alphavir.  The small blue circles indicate YSOs from infrared surveys \citep{Whitney08,GC09,Seale14} and the large green circles indicate HII regions and their sizes from \cite{Henize56}. The black lines show the boundaries used to split the Ridge into four different regions to analyze and the dashed lines show the observation footprint. }
    \label{fig:alphavir map}
\end{figure}

To further investigate the physical conditions in the Ridge, we next look for variation in \alphavir values and other physical properties within the region. Figure\,\ref{fig:alphavir map} shows a map of the Ridge colored by the \alphavir parameter measured for the structure. In this map, we use structures found with \texttt{quickclump} to minimize overlap in the map. Where there is overlap along the line of sight, the average is shown. Also shown in the map are YSOs from \cite{Whitney08}, \cite{GC09}, and \cite{Seale14}, as well as HII regions and their sizes from \cite{Henize56}. There appears to be some regions that tend toward higher or lower \alphavir parameters, but there does not seem to be any consistent trend north-to-south along the Ridge.

We split the structures in the Ridge into four main regions - {a Southern region, then an Eastern, a Central, and a Western region to the north }(hereafter referred to as NE, NC, and NW, respectively). The {boundaries} between these regions are shown in Figure\,\ref{fig:alphavir map}. There is one major clump not included in any region, just below the borders of the NC and NW regions. This clump is partly cut off by the edge of the observation footprint and so does not have a reliable \alphavir {measurement. It is left out of the analysis.}

\begin{figure*}
    \centering
    \includegraphics[width=0.8\textwidth]{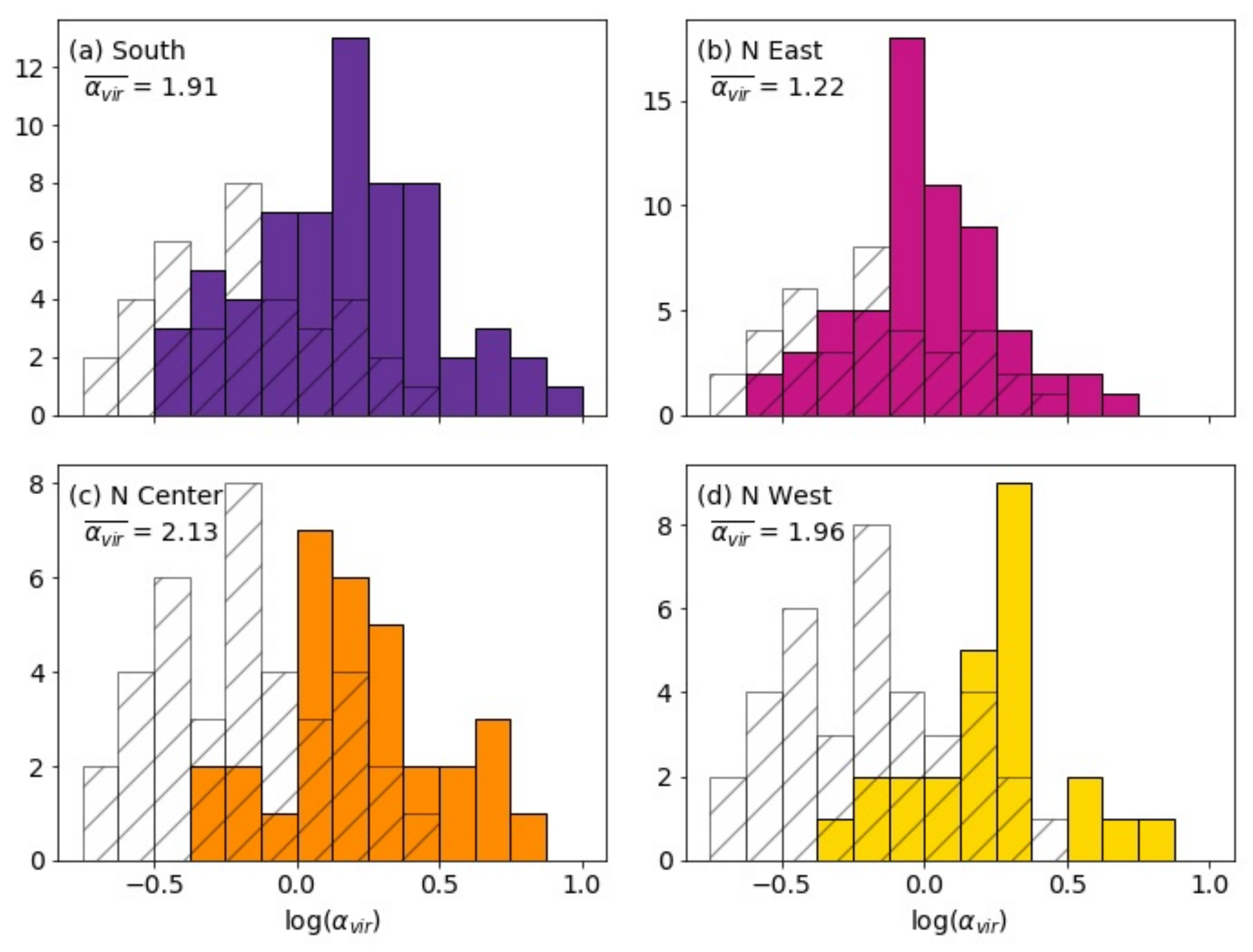}
    \caption{Histograms of the \alphavir parameter by region in the Ridge, {derived from \thirtCO(1-0)} (Figure\,\ref{fig:alphavir map} shows {the extent of the regions}). Overplotted in black hatching is the \alphavir distribution of 30~Dor. The average \alphavir parameter is shown in the top left of the panel for each region. A series of K-S {tests} indicates that the NE region (pink) has a different distribution than each of the other regions in the Ridge. The NE region tends towards lower values of \alphavir than the rest of the Ridge. }
    \label{fig:alpha region hists}
\end{figure*}

To compare the \alphavir values in each of the four regions, we plot histograms of their values in Figure\,\ref{fig:alpha region hists}. The shape of these distributions appear by eye primarily Gaussian. We perform a K-S test on each pairing of regions to determine if their values of \alphavir are drawn from a common distribution. The resulting p-values of the K-S test for each pairing are presented in Table\,\ref{tab:KS alphavir}, where a p-value of less than 0.05 means we reject the null hypothesis (that the \alphavir values for the two regions are drawn from the same distribution) with at least 95\% confidence.

\begin{table}
    \centering
    \caption{K-S test p-values of \alphavir distributions in pairings of regions within the Ridge, {derived from \thirtCO(1-0).}}
    \begin{tabular}{c|c|c|c}
        \hline
        \hline
          & N East & N Center & N West \\
         \hline 
         South  & 0.016 & 0.702 & 0.305 \\
         N East & - & 0.002 & 0.001 \\
         N Center & - & - & 0.751 \\
    \end{tabular}
    \label{tab:KS alphavir}
\end{table}

The South, NC, and NW regions are all consistent with being drawn from the same distribution. The NE region, however, has {K-S test p-values less than 0.05} when paired with each of the other regions, suggesting that its \alphavir values are not drawn from the same distribution. This suggests that there is a difference in the physical conditions in the NE region that is causing lower values of \alphavir in that region compared to the rest of the Ridge.

\begin{figure}
    \centering
    \includegraphics[width=0.45\textwidth]{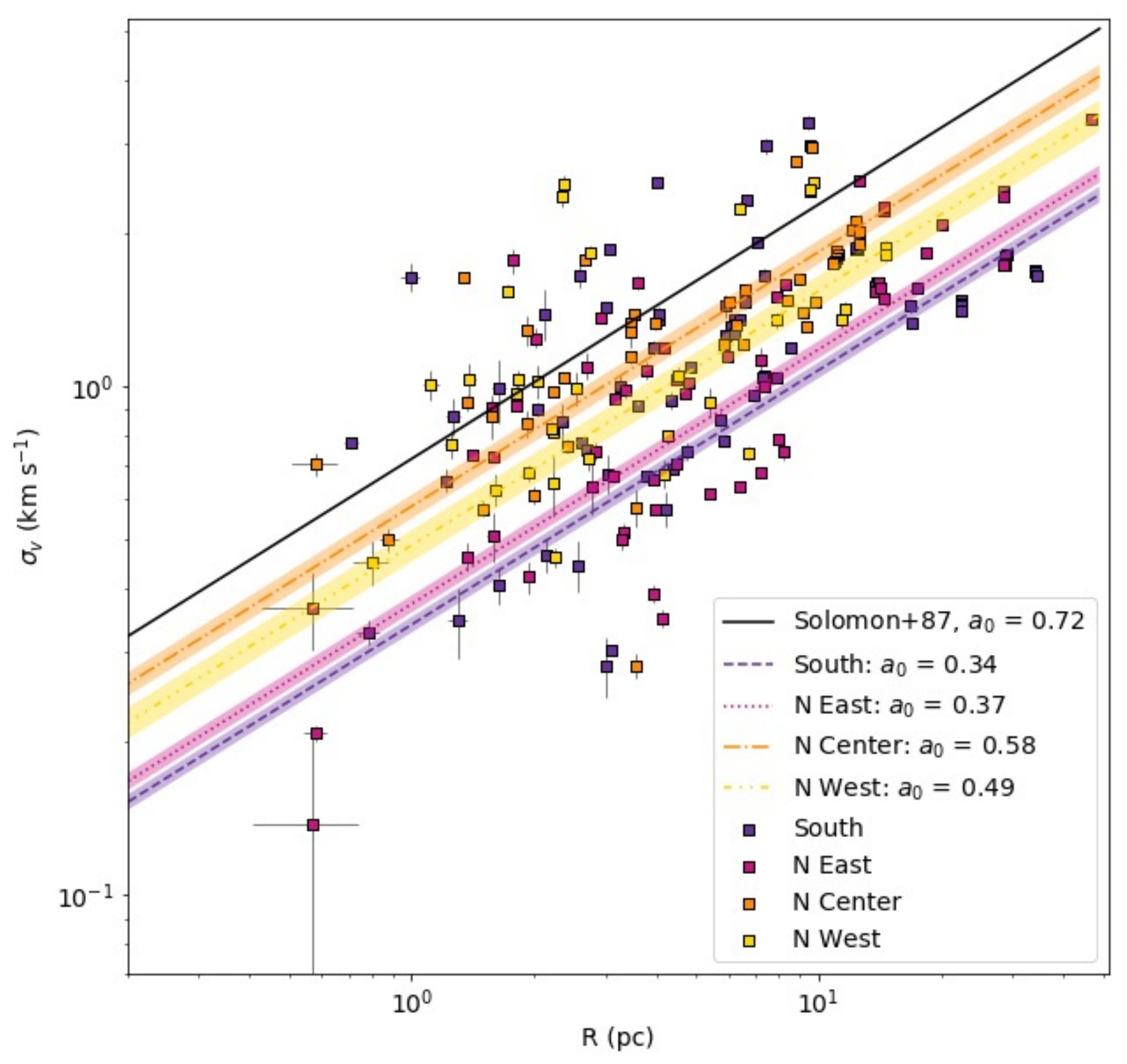}
    \caption{Linewidths plotted against sizes of structures in the Ridge, colored by their region within the Ridge. We fit a power law with a fixed slope of $a_1=0.5$ to each region's structures and find that the South and NE regions have the lowest intercepts, indicating low kinetic energy in these regions. The NC region has the highest intercept, indicating that it has the highest kinetic energy. }
    \label{fig:regions SL}
\end{figure}

We look into this difference further by comparing the size-linewidth relation in these different regions within the Ridge. Similar to \S\ref{sec:SL plots}, we fit the power law relation between $\sigma_v$ and $R$ for the structures in each of these regions with the slope of the power law fixed at $a_1 = 0.5$ (Figure\,\ref{fig:regions SL}). For these regions, we fit $a_0 = 0.33\pm0.01$, $0.35\pm0.01$, $0.55\pm0.03$, and $0.46\pm0.03$ for the South, NE, NC, and NW regions, respectively. This indicates that the South and NE regions have similarly low kinetic energies, while the NW region has higher kinetic energy and the NC has the highest kinetic energy. This makes sense since the NC region hosts the large HII region, N171 \citep{Henize56}.

The low kinetic energy in the NE region relative to the NC and NW {regions} could explain why its \alphavir values are lower. However, the kinetic energy in the {Southern} region appears nearly as low and its \alphavir values are comparable with those in the NC and NW regions. 
To explain these variations in \alphavir values, we also consider maps of \nh in the Ridge based on fitting \radex models to low resolution \twelveCO\ and \thirtCO\ emission in \cite{Finn21}. This map indicates that the NE region has higher gas densities than the {Southern} region, which would result in the NE having lower \alphavir values despite having comparably low kinetic energy. It is unclear from these plots what might be causing the higher densities in the NE region.

\section{Spatial Dependence} \label{sec: spatial dependence}

\begin{figure}
    \centering
    \includegraphics[width=0.45\textwidth]{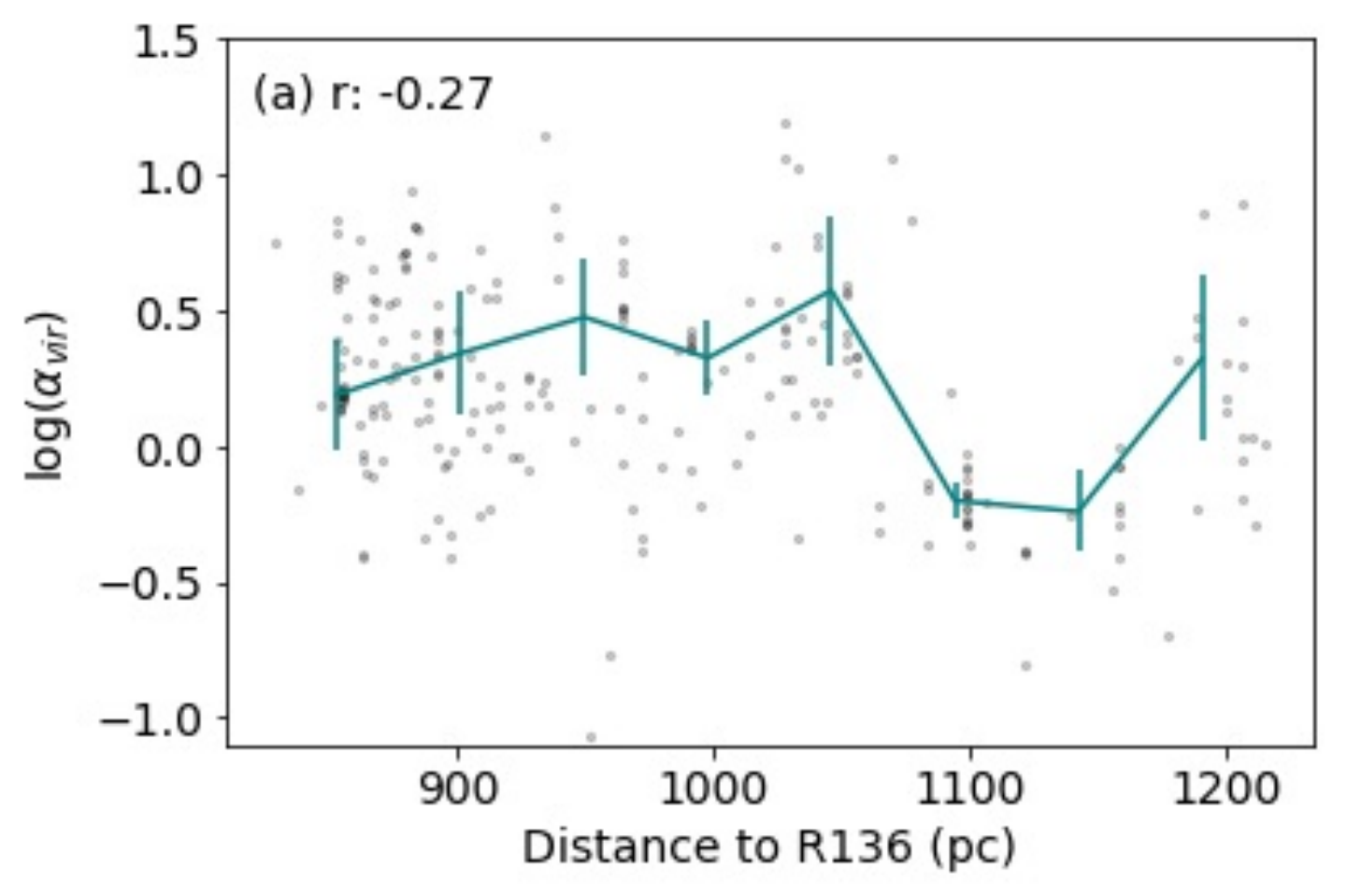}
    \includegraphics[width=0.45\textwidth]{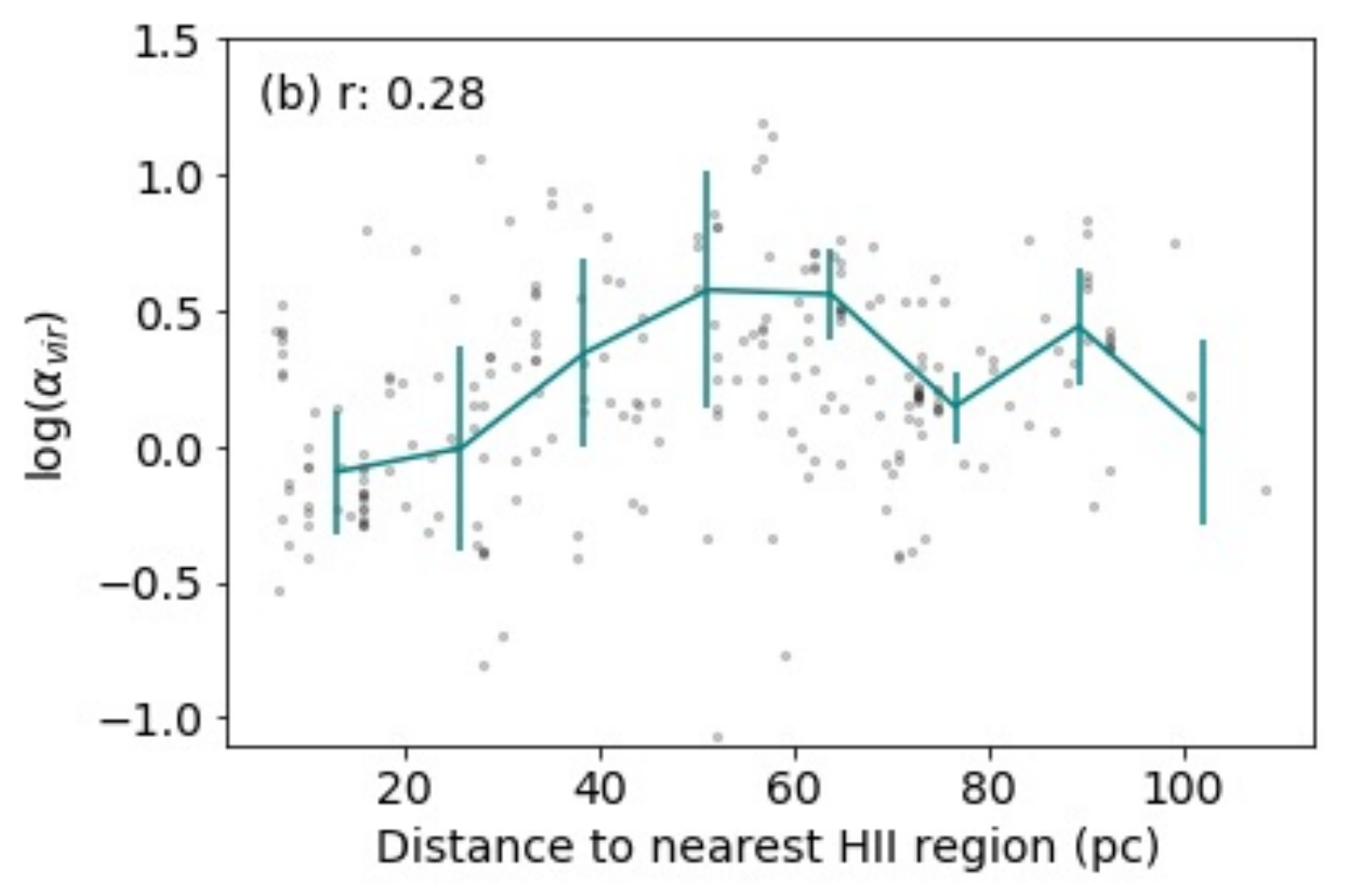}
    \includegraphics[width=0.45\textwidth]{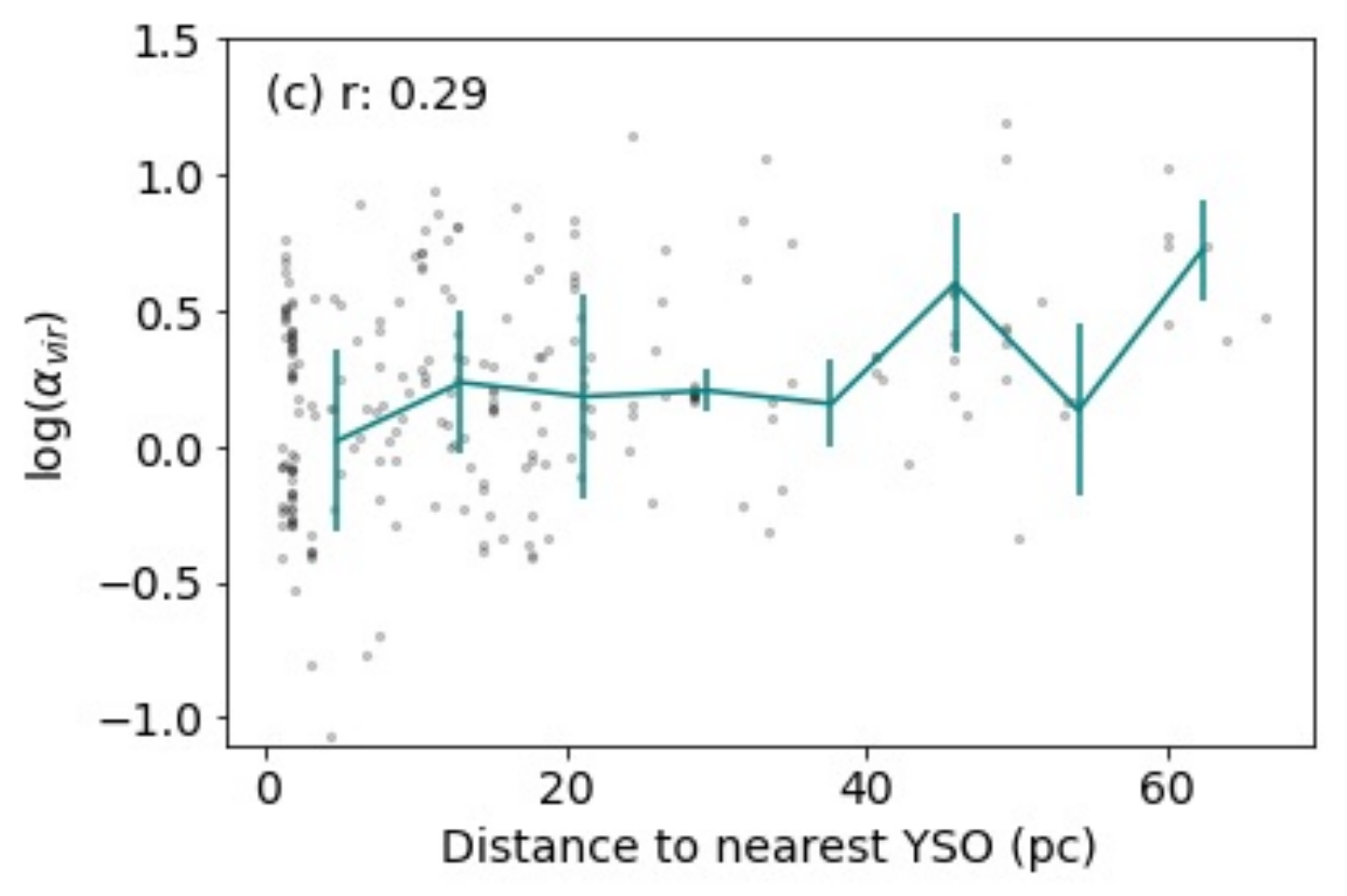}
    \caption{Values of \alphavir in the Ridge plotted against the structure's distance to R136 (\emph{top}), the nearest HII region (\emph{middle}), and the nearest YSO (\emph{bottom}). All three show similarly weak correlations based on Pearson's correlation coefficients, shown in the top left corner. }
    \label{fig:alphavir dist}
\end{figure}

To investigate what is driving the variations in \alphavir values in the Ridge, we consider how \alphavir varies with proximity to the nearest YSO, the nearest HII region, and to R136 in 30~Dor. These plots are shown in Figure\,\ref{fig:alphavir dist} and for each we calculate a Pearson's correlation coefficient, $r$.

All three of these show similarly weak correlations. The variation of \alphavir with distance to R136 shows the weakest correlation, with a coefficient of -0.27. This agrees with our by-eye assessment of Figure\,\ref{fig:alphavir map} that there appears to be little cohesive north-to-south trend across the Ridge. This result suggests that the super star cluster R136 and in general that massive star formation to the north is not strongly affecting the physical conditions in the Ridge at this distance. This is consistent with the findings of \cite{Wong22} that there is little correlation with \alphavir for clouds in 30 Dor and their distance to R136.

The weak correlation in the figure appears most dominated by lower \alphavir values around a distance of 1.1 kpc from R136, which corresponds to a declination of approximately $-70^\circ 26.4'$. This falls halfway along the South region defined in Figure\,\ref{fig:alphavir map}, and is near several YSOs and an HII region.

The correlation coefficient between \alphavir and distance to the nearest YSO is 0.29, and between \alphavir and distance to the nearest HII region is 0.28. The weak correlation suggests that local influences such as radiation and thermal pressure from nearby star formation have a small effect on molecular gas conditions. Structures that have a nearby YSO or HII region tend to have lower \alphavir. This could be because the YSOs and HII regions are creating gas conditions that are conducive for further star formation, or because they {are} more likely to be associated with gas that has the right conditions for star formation.

\section{Discussion} \label{sec:discussion}

With this analysis, we hope to better understand what is driving the differences in star formation activity between the Molecular Ridge and {the} nearby massive-star-forming regions 30~Dor, N159, and N113. By comparing the fractal dimensions of 30~Dor and the Ridge, it seems that the hierarchical morphology in the two regions is similar and thus the state of fragmentation is not driving the difference in star formation.

From Figure\,\ref{fig:SL slope}, we conclude that the Ridge has significantly lower kinetic energy per size scale than the massive star forming regions, while Figure\,\ref{fig:alphavir hist} demonstrates that despite the low kinetic energy, the Ridge still has higher values of \alphavir on average. This must be driven by relatively low surface densities in the Ridge, which is confirmed by Figure\,\ref{fig:Sigma hist}. All of this evidence suggests that the Ridge's lack of massive star formation is likely driven by a paucity of dense gas relative to regions like 30~Dor, rather than collapse being suppressed by excess turbulence in the Ridge. However, as \cite{Finn21} point out, surface density cannot account for differences in line-of-sight length and cannot directly trace the true volume density of the gas. To directly compare the volume density of the gas in the Ridge and the massive star forming regions, we would need to observe multiple \twelveCO\ and \thirtCO\ emission lines in other regions and perform the same \radex fitting done in \cite{Finn21}.

As shown in Figure\,\ref{fig:virial}, the systematic uncertainty in the mass estimates makes it difficult to discuss the absolute position of the structures relative to virial equilibrium. The Ridge could have a higher distribution of \alphavir values for a few different reasons: the clouds are subjected to an external pressure, the clouds are in free fall, or the clouds are unbounded. The even distribution of the Ridge clouds along the virial line in Figure\,\ref{fig:virial} rather than clustering around a consistent range of pressures makes it seem unlikely that an external pressure has a large influence here. The clouds being in free fall collapse could also explain the higher \alphavir values, but this scenario seems inconsistent with the low kinetic energies indicated in the size-linewidth plots (Figure\,\ref{fig:SL intercept}) and the low amount of star formation. It seems most likely that the clouds in the Ridge tend to be less gravitationally bound than the clouds in the other star forming regions. This could be due to low gas density, which would make it harder for the clouds to hold together or begin gravitational collapse. 

This finding may also have implications for the types of stars and clusters formed based on the density of gas forming those stars.  \cite{Indebetouw08} find that the Ridge is still forming stars, but preferentially forms low-mass clusters that do not sample a standard initial mass function \citep[IMF; such as ][]{Kroupa02}  well and so do not produce many massive stars. This aligns well with our finding that at this resolution, structures in the Ridge have surface densities that fall below the threshold for massive star formation of $A_V>8$ {mag} from \cite{Lada10}. Meanwhile, 30~Dor structures are distributed around that threshold, and {the} massive star forming regions N159 and N113 have average surface densities well above that threshold.

The gradient in star forming potential from 30~Dor, N159, and through the Ridge seems consistent with the interaction histories of the LMC and SMC proposed by \cite{Besla12} in which the SMC collided with the LMC in the past 250 Myr. Recent numerical simulations have further constrained that interaction to the SMC colliding with the LMC 140-160 Myr ago with an impact parameter of $\sim5$ kpc \citep{Choi22}. {The Magellanic Bridge connecting the two galaxies is likely the result of such an interaction, and the Molecular Ridge extends from 30~Dor in the direction of the Bridge, suggesting a potential connection between the two structures.} The current relative velocity of the LMC and SMC are estimated to be $\sim100$ km s$^{-1}$ \citep{Zivick19}, meaning that during such a collision, the SMC's motion through the LMC would be supersonic for the molecular gas, creating shocks and over-densities in the molecular clouds. {The extended Magellanic Streams seen in HI gas are also likely caused in part by close interactions between the LMC and SMC, as well as their interactions with the Milky Way \citep[e.g.,  ][]{Lucchini21}, although this structure occurs on much larger size scales than the regions studied here.}

This interaction scenario is also supported by the findings of \cite{Furuta19} that the gas in the regions around 30~Dor and N159 has a lower measured $A_V/N(\text{H})$ than the rest of the LMC and is instead consistent with gas in the SMC. \cite{Furuta21} proposed a geometry in which gas from the SMC is colliding with the LMC disk moving north to south, forming R136 in 30~Dor {and} then N159.

The LMC-SMC interaction and subsequent tidal effects would have increased the turbulent motion in the molecular gas, which could have led to the formation of R136 as well as the older populations in the 30~Dor region \citep{Rahner18}. To create an SSC like R136 that has a mass of $\sim10^5$ M$_\odot$, the initial molecular cloud would have needed to start with a mass of at least $\sim2\times10^5$ M$_\odot$, assuming a maximum star formation efficiency for SSCs of 50\% \citep{AshmanZepf01,Kroupa01,Grudic18}. \cite{Johnson15} further required that to form an SSC, the initial molecular cloud must contain this large mass within a maximum radius of 25 pc.
None of the molecular cloud structures in 30~Dor meet these criteria, and so are not capable of creating an SSC like R136, suggesting that the gas conditions during the peak of star formation in 30~Dor were much more extreme than they currently are. The molecular cloud that created R136 more likely appeared similar in physical conditions to the potential SSC precursor cloud observed in the merging Antennae galaxies \citep{Johnson15}. \cite{Finn2019} observed that this cloud has a large \alphavir parameter and would require high external pressure in order to remain bound. They also found evidence that this pressure is supplied by cloud-cloud collision. \cite{Fukui17} presented evidence that R136 was also created by a tidally-induced collision of large scale HI clouds, which would align with the idea that the molecular cloud precursor to R136 was subject to high external pressure from a galactic interaction.

\cite{Fukui2015} also found evidence {for a cloud-cloud collision} triggering the formation of high-mass YSOs in N159, and follow-up work by \cite{Fukui19} and \cite{Tokuda19} suggests that this collision is associated with the same large scale colliding flows that triggered the formation of R136 cited above. N159 is currently forming more massive star clusters than the 30~Dor region and hosts a massive molecular core of $\sim10^4$ M$_\odot$ within a $\sim1$ pc radius \citep{Tokuda22}, but it does not have any molecular cloud structures measured in this analysis that meet the SSC-forming criteria cited above \citep[a mass of at least $2\times10^5$ M$_\odot$ within a radius of 25 pc; ][]{Johnson15}. This ongoing star formation in N159 is consistent with our finding that the structures are currently near or below virial equilibrium, and so are likely to collapse and form stars, but they do not show signs of requiring a high external pressure to be bound as seen in SSC-forming clouds. 

If 30~Dor and N159 experienced cloud collisions induced by the interaction of the SMC and LMC, it would be reasonable to expect that as the SMC moved away from the LMC, the gas conditions became less extreme (and so N159 is not forming stars as intensely as 30~Dor once was) and some molecular gas may also have been pulled out of the SMC in the direction of the Magellanic Bridge, {meaning that the Ridge may even be a denser extension of the Magellanic Bridge}. The Ridge could be this gas, and our finding that it is less dense than the gas in 30~Dor and N159 is consistent with that interpretation. It is also consistent with our finding that the northern regions have higher kinetic energy and densities and so are more similar to 30~Dor and N159 than the molecular gas towards the south of the Ridge.

This proposed scenario demonstrates how galaxy interactions can create regions that both over- and under-produce stars when compared to often-applied scaling relations. This effect can be important when accounting for star formation in galaxy simulations, especially given the importance of dwarf galaxy mergers in the evolution of galaxies over cosmic time. It would be interesting to see if any simulations of dwarf galaxy interactions,  {especially of the SMC and LMC interaction specifically, are able to recreate the morphology of the Ridge and the gas conditions that we see along its extent.} Such simulations may also help clarify the timescales of the interaction between the galaxies and the subsequent tidal effects that may lead to further cloud collisions and the eventual onset of star formation in regions like 30~Dor.

\section{Conclusions} \label{sec:conclusions}

We present a comparison of \thirtCO\ observations of the Molecular Ridge, 30~Dor, N159, and N113 in the LMC. The latter three regions are all actively forming massive stars, while the Ridge is not, despite its large reservoir of molecular gas. We use dendrograms and clump-finding algorithms to segment the emission and analyze the physical conditions of those structures. Our major findings are summarized below.

\begin{itemize}

    \item The Ridge, 30~Dor, and N113 have fractal dimensions of $D_2 = 1.50\pm0.02$, $1.44\pm0.02$, $1.42\pm0.07$ respectively. These are similar enough that it seems unlikely the difference in star formation activity between the regions is related to a difference in cloud morphology and hierarchical structure. These values are also consistent with other measurements of the fractal dimension within the LMC, SMC, and Milky Way. {(\S\ref{subsec: fractal dimension})}
    
    \item Comparing size-linewidth relations in the Ridge, 30~Dor, N159, and N113 indicates that the Ridge has significantly lower kinetic energy at given size scales when compared to the massive star forming regions. This would rule out the possibility that the Ridge has lower rates of massive star formation because it is supported against collapse due to excess kinetic energy. {(\S\ref{sec:SL plots})}
    
    \item The Ridge has higher values of \alphavir than 30~Dor, although the absolute scaling of these values relative to virial equilibrium is unclear due to large uncertainties in the mass estimate. This appears to be driven by significantly lower surface densities in the Ridge, since we know from the size-linewidth relations that the Ridge also has lower kinetic energy than 30~Dor. We find as well that the structures in the Ridge fall below the \cite{Lada10} threshold of $A_V>8$ {mag} for massive star formation, while 30~Dor structures are distributed around this threshold and the average surface densities in N159 and N113 are well above this threshold. These results suggest that the Ridge has lower rates of massive star formation because it has significantly less dense gas than 30~Dor, although from \cite{Finn21} we know that the surface density of the gas does not necessarily trace the non-LTE-fitted volume density of the gas. {(\S\ref{sec: virial plot})}
    
    \item Within the Ridge, there is some variation in physical properties. The region in the northeast has a significantly lower \alphavir distribution than the other regions, likely partially driven by having a low specific kinetic energy and relatively high densities. {(\S\ref{sec: ridge variation})}
    
    \item The \alphavir values measured for structures are only weakly correlated with distance from the super star cluster R136 in 30~Dor, suggesting that such star clusters do not affect gas properties at kiloparsec distances. The \alphavir values similarly show only a weak correlation with the distances to the nearest YSO and the nearest HII region, suggesting that local star formation has a minimal influence on gas conditions or that the YSOs are slightly more likely to be spatially coincident with gas that has the right conditions for further star formation. {(\S\ref{sec: spatial dependence})}
    
\end{itemize}

\begin{acknowledgements}

This research is supported by NSF grants 1413231 and 1716335 (PI: K.~Johnson) and NSF grant 2009624 and NSF AAG award AST 1312902 to U. Virginia (PI: R.~Indebetouw). This material is based upon work supported by the National Science Foundation Graduate Research Fellowship Program under Grant No. 1842490. Any opinions, findings, and conclusions or recommendations expressed in this material are those of the author(s) and do not necessarily reflect the views of the National Science Foundation. T.W. acknowledges support from NSF AAG award 2009849. 
The material is based upon work supported by NASA under award number 80GSFC21M0002 (M.S.). 
K.T. acknowledges support from NAOJ ALMA Scientific Research grant Nos. 2022-22B, and Grants-in-Aid for Scientific Research (KAKENHI) of Japan Society for the Promotion of Science (JSPS; grant Nos., JP21H00049, and JP21K13962).

This paper makes use of the following ALMA data: ADS/JAO.ALMA\#2012.1.00554.S, \\ADS/JAO.ALMA\#2015.1.00196.S, \\ADS/JAO.ALMA\#2015.1.01388.S, \\ADS/JAO.ALMA\#2017.1.00271.S, \\ADS/JAO.ALMA\#2019.1.00843. ALMA is a partnership of ESO (representing its member states), NSF (USA) and NINS (Japan), together with NRC (Canada), NSC and ASIAA (Taiwan), and KASI (Republic of Korea), in cooperation with the Republic of Chile. The Joint ALMA Observatory is operated by ESO, AUI/NRAO and NAOJ. The National Radio Astronomy Observatory is a facility of the National Science Foundation operated under cooperative agreement by Associated Universities, Inc.

Based on observations with the Atacama Pathfinder EXperiment (APEX) telescope. APEX is a collaboration between the Max Planck Institute for Radio Astronomy, the European Southern Observatory, and the Onsala Space Observatory. Swedish observations on APEX are supported through Swedish Research Council grant No 2017-00648.

\facility{ALMA}

\software{Pipeline-CASA51-P2-B v.40896 \citep{Davis21}, 
    CASA \citep[v.5.1.1-5, v.5.6.1; ][]{McMullin07}, 
    \texttt{astrodendro} \citep{Rosolowsky08}, 
    \texttt{quickclump} \citep{Sidorin17}, 
    Astropy \citep{astropy}, 
    Matplotlib \citep{matplotlib}, 
    NumPy \citep{numpy}, 
    SciPy \citep{scipy}
}

\end{acknowledgements}

\bibliographystyle{yahapj}
\bibliography{references.bib}

\begin{thebibliography}{}
\providecommand\natexlab[1]{#1}
\providecommand\JournalTitle[1]{#1}

\bibitem[{{Ashman} \& {Zepf}(2001)}]{AshmanZepf01}
{Ashman}, K.~M., \& {Zepf}, S.~E. 2001,
  \href{http://dx.doi.org/10.1086/323133}{\JournalTitle{\aj}, 122, 1888}

\bibitem[{{Astropy Collaboration} {et~al.}(2013){Astropy Collaboration},
  {Robitaille}, {Tollerud}, {Greenfield}, {Droettboom}, {Bray}, {Aldcroft},
  {Davis}, {Ginsburg}, {Price-Whelan}, {Kerzendorf}, {Conley}, {Crighton},
  {Barbary}, {Muna}, {Ferguson}, {Grollier}, {Parikh}, {Nair}, {Unther},
  {Deil}, {Woillez}, {Conseil}, {Kramer}, {Turner}, {Singer}, {Fox}, {Weaver},
  {Zabalza}, {Edwards}, {Azalee Bostroem}, {Burke}, {Casey}, {Crawford},
  {Dencheva}, {Ely}, {Jenness}, {Labrie}, {Lim}, {Pierfederici}, {Pontzen},
  {Ptak}, {Refsdal}, {Servillat}, \& {Streicher}}]{astropy}
{Astropy Collaboration}, {Robitaille}, T.~P., {Tollerud}, E.~J., {et~al.} 2013,
  \href{http://dx.doi.org/10.1051/0004-6361/201322068}{\JournalTitle{\aap},
  558, A33}

\bibitem[{{Besla} {et~al.}(2012){Besla}, {Kallivayalil}, {Hernquist}, {van der
  Marel}, {Cox}, \& {Kere{\v{s}}}}]{Besla12}
{Besla}, G., {Kallivayalil}, N., {Hernquist}, L., {et~al.} 2012,
  \href{http://dx.doi.org/10.1111/j.1365-2966.2012.20466.x}{\JournalTitle{\mnras},
  421, 2109}

\bibitem[{{Bica} {et~al.}(1996){Bica}, {Claria}, {Dottori}, {Santos}, \&
  {Piatti}}]{Bica96}
{Bica}, E., {Claria}, J.~J., {Dottori}, H., {Santos}, J.~F.~C., J., \&
  {Piatti}, A.~E. 1996,
  \href{http://dx.doi.org/10.1086/192251}{\JournalTitle{\apjs}, 102, 57}

\bibitem[{{Bolatto} {et~al.}(2013){Bolatto}, {Wolfire}, \& {Leroy}}]{Bolatto13}
{Bolatto}, A.~D., {Wolfire}, M., \& {Leroy}, A.~K. 2013,
  \href{http://dx.doi.org/10.1146/annurev-astro-082812-140944}{\JournalTitle{\araa},
  51, 207}

\bibitem[{{Calzetti} {et~al.}(2007){Calzetti}, {Kennicutt}, {Engelbracht},
  {Leitherer}, {Draine}, {Kewley}, {Moustakas}, {Sosey}, {Dale}, {Gordon},
  {Helou}, {Hollenbach}, {Armus}, {Bendo}, {Bot}, {Buckalew}, {Jarrett}, {Li},
  {Meyer}, {Murphy}, {Prescott}, {Regan}, {Rieke}, {Roussel}, {Sheth}, {Smith},
  {Thornley}, \& {Walter}}]{Calzetti07}
{Calzetti}, D., {Kennicutt}, R.~C., {Engelbracht}, C.~W., {et~al.} 2007,
  \href{http://dx.doi.org/10.1086/520082}{\JournalTitle{\apj}, 666, 870}

\bibitem[{{Chen} {et~al.}(2010){Chen}, {Indebetouw}, {Chu}, {Gruendl},
  {Testor}, {Heitsch}, {Seale}, {Meixner}, \& {Sewilo}}]{Chen10}
{Chen}, C.~H.~R., {Indebetouw}, R., {Chu}, Y.-H., {et~al.} 2010,
  \href{http://dx.doi.org/10.1088/0004-637X/721/2/1206}{\JournalTitle{\apj},
  721, 1206}

\bibitem[{{Chevance} {et~al.}(2020){Chevance}, {Madden}, {Fischer}, {Vacca},
  {Lebouteiller}, {Fadda}, {Galliano}, {Indebetouw}, {Kruijssen}, {Lee},
  {Poglitsch}, {Polles}, {Cormier}, {Hony}, {Iserlohe}, {Krabbe}, {Meixner},
  {Sabbi}, \& {Zinnecker}}]{Chevance20}
{Chevance}, M., {Madden}, S.~C., {Fischer}, C., {et~al.} 2020,
  \href{http://dx.doi.org/10.1093/mnras/staa1106}{\JournalTitle{\mnras}, 494,
  5279}

\bibitem[{{Choi} {et~al.}(2022){Choi}, {Olsen}, {Besla}, {van der Marel},
  {Zivick}, {Kallivayalil}, \& {Nidever}}]{Choi22}
{Choi}, Y., {Olsen}, K. A.~G., {Besla}, G., {et~al.} 2022,
  \href{http://dx.doi.org/10.3847/1538-4357/ac4e90}{\JournalTitle{\apj}, 927,
  153}

\bibitem[{{Cignoni} {et~al.}(2015){Cignoni}, {Sabbi}, {van der Marel}, {Tosi},
  {Zaritsky}, {Anderson}, {Lennon}, {Aloisi}, {de Marchi}, {Gouliermis},
  {Grebel}, {Smith}, \& {Zeidler}}]{Cignoni15}
{Cignoni}, M., {Sabbi}, E., {van der Marel}, R.~P., {et~al.} 2015,
  \href{http://dx.doi.org/10.1088/0004-637X/811/2/76}{\JournalTitle{\apj}, 811,
  76}

\bibitem[{{Cohen} {et~al.}(1988){Cohen}, {Dame}, {Garay}, {Montani}, {Rubio},
  \& {Thaddeus}}]{Cohen88}
{Cohen}, R.~S., {Dame}, T.~M., {Garay}, G., {et~al.} 1988,
  \href{http://dx.doi.org/10.1086/185243}{\JournalTitle{\apjl}, 331, L95}

\bibitem[{{Davies} {et~al.}(1976){Davies}, {Elliott}, \& {Meaburn}}]{Davies76}
{Davies}, R.~D., {Elliott}, K.~H., \& {Meaburn}, J. 1976,
  \JournalTitle{\memras}, 81, 89

\bibitem[{{Davis}(2021)}]{Davis21}
{Davis}. 2021, \JournalTitle{in prep}

\bibitem[{{Dobashi} {et~al.}(2008){Dobashi}, {Bernard}, {Hughes}, {Paradis},
  {Reach}, \& {Kawamura}}]{Dobashi08}
{Dobashi}, K., {Bernard}, J.~P., {Hughes}, A., {et~al.} 2008,
  \href{http://dx.doi.org/10.1051/0004-6361:20079151}{\JournalTitle{\aap}, 484,
  205}

\bibitem[{{Elmegreen} \& {Falgarone}(1996)}]{ElmegreenFalgarone96}
{Elmegreen}, B.~G., \& {Falgarone}, E. 1996,
  \href{http://dx.doi.org/10.1086/178009}{\JournalTitle{\apj}, 471, 816}

\bibitem[{{Elmegreen} {et~al.}(2001){Elmegreen}, {Kim}, \&
  {Staveley-Smith}}]{Elmegreen01}
{Elmegreen}, B.~G., {Kim}, S., \& {Staveley-Smith}, L. 2001,
  \href{http://dx.doi.org/10.1086/319021}{\JournalTitle{\apj}, 548, 749}

\bibitem[{{Falgarone} {et~al.}(1991){Falgarone}, {Phillips}, \&
  {Walker}}]{Falgarone91}
{Falgarone}, E., {Phillips}, T.~G., \& {Walker}, C.~K. 1991,
  \href{http://dx.doi.org/10.1086/170419}{\JournalTitle{\apj}, 378, 186}

\bibitem[{{Field} {et~al.}(2011){Field}, {Blackman}, \& {Keto}}]{Field11}
{Field}, G.~B., {Blackman}, E.~G., \& {Keto}, E.~R. 2011,
  \href{http://dx.doi.org/10.1111/j.1365-2966.2011.19091.x}{\JournalTitle{\mnras},
  416, 710}

\bibitem[{{Finn} {et~al.}(2019){Finn}, {Johnson}, {Brogan}, {Wilson},
  {Indebetouw}, {Harris}, {Kamenetzky}, \& {Bemis}}]{Finn2019}
{Finn}, M.~K., {Johnson}, K.~E., {Brogan}, C.~L., {et~al.} 2019,
  \href{http://dx.doi.org/10.3847/1538-4357/ab0d1e}{\JournalTitle{\apj}, 874,
  120}

\bibitem[{{Finn} {et~al.}(2021){Finn}, {Indebetouw}, {Johnson}, {Costa},
  {Chen}, {Kawamura}, {Onishi}, {Ott}, {Tokuda}, {Wong}, \&
  {Zahorecz}}]{Finn21}
{Finn}, M.~K., {Indebetouw}, R., {Johnson}, K.~E., {et~al.} 2021,
  \href{http://dx.doi.org/10.3847/1538-4357/ac090c}{\JournalTitle{\apj}, 917,
  106}

\bibitem[{{Fomalont} {et~al.}(2014){Fomalont}, {van Kempen}, {Kneissl},
  {Marcelino}, {Barkats}, {Corder}, {Cortes}, {Hills}, {Lucas}, {Manning}, \&
  {Peck}}]{Fomalont14}
{Fomalont}, E., {van Kempen}, T., {Kneissl}, R., {et~al.} 2014,
  \JournalTitle{The Messenger}, 155, 19

\bibitem[{{Fukui} {et~al.}(2017){Fukui}, {Tsuge}, {Sano}, {Bekki}, {Yozin},
  {Tachihara}, \& {Inoue}}]{Fukui17}
{Fukui}, Y., {Tsuge}, K., {Sano}, H., {et~al.} 2017,
  \href{http://dx.doi.org/10.1093/pasj/psx032}{\JournalTitle{\pasj}, 69, L5}

\bibitem[{{Fukui} {et~al.}(2008){Fukui}, {Kawamura}, {Minamidani}, {Mizuno},
  {Kanai}, {Mizuno}, {Onishi}, {Yonekura}, {Mizuno}, {Ogawa}, \&
  {Rubio}}]{NANTEN}
{Fukui}, Y., {Kawamura}, A., {Minamidani}, T., {et~al.} 2008,
  \href{http://dx.doi.org/10.1086/589833}{\JournalTitle{\apjs}, 178, 56}

\bibitem[{{Fukui} {et~al.}(2015){Fukui}, {Harada}, {Tokuda}, {Morioka},
  {Onishi}, {Torii}, {Ohama}, {Hattori}, {Nayak}, {Meixner}, {Sewi{\l}o},
  {Indebetouw}, {Kawamura}, {Saigo}, {Yamamoto}, {Tachihara}, {Minamidani},
  {Inoue}, {Madden}, {Galametz}, {Lebouteiller}, {Mizuno}, \&
  {Chen}}]{Fukui2015}
{Fukui}, Y., {Harada}, R., {Tokuda}, K., {et~al.} 2015,
  \href{http://dx.doi.org/10.1088/2041-8205/807/1/L4}{\JournalTitle{\apjl},
  807, L4}

\bibitem[{{Fukui} {et~al.}(2019){Fukui}, {Tokuda}, {Saigo}, {Harada},
  {Tachihara}, {Tsuge}, {Inoue}, {Torii}, {Nishimura}, {Zahorecz}, {Nayak},
  {Meixner}, {Minamidani}, {Kawamura}, {Mizuno}, {Indebetouw}, {Sewi{\l}o},
  {Madden}, {Galametz}, {Lebouteiller}, {Chen}, \& {Onishi}}]{Fukui19}
{Fukui}, Y., {Tokuda}, K., {Saigo}, K., {et~al.} 2019,
  \href{http://dx.doi.org/10.3847/1538-4357/ab4900}{\JournalTitle{\apj}, 886,
  14}

\bibitem[{{Furuta} {et~al.}(2019){Furuta}, {Kaneda}, {Kokusho}, {Ishihara},
  {Nakajima}, {Fukui}, \& {Tsuge}}]{Furuta19}
{Furuta}, T., {Kaneda}, H., {Kokusho}, T., {et~al.} 2019,
  \href{http://dx.doi.org/10.1093/pasj/psz078}{\JournalTitle{\pasj}, 71, 95}

\bibitem[{{Furuta} {et~al.}(2021){Furuta}, {Kaneda}, {Kokusho}, {Nakajima},
  {Fukui}, \& {Tsuge}}]{Furuta21}
---. 2021, \href{http://dx.doi.org/10.1093/pasj/psab052}{\JournalTitle{\pasj},
  73, 864}

\bibitem[{{Goodman} {et~al.}(1998){Goodman}, {Barranco}, {Wilner}, \&
  {Heyer}}]{Goodman98}
{Goodman}, A.~A., {Barranco}, J.~A., {Wilner}, D.~J., \& {Heyer}, M.~H. 1998,
  \href{http://dx.doi.org/10.1086/306045}{\JournalTitle{\apj}, 504, 223}

\bibitem[{{Grudi{\'c}} {et~al.}(2018){Grudi{\'c}}, {Hopkins},
  {Faucher-Gigu{\`e}re}, {Quataert}, {Murray}, \& {Kere{\v{s}}}}]{Grudic18}
{Grudi{\'c}}, M.~Y., {Hopkins}, P.~F., {Faucher-Gigu{\`e}re}, C.-A., {et~al.}
  2018, \href{http://dx.doi.org/10.1093/mnras/sty035}{\JournalTitle{\mnras},
  475, 3511}

\bibitem[{{Gruendl} \& {Chu}(2009)}]{GC09}
{Gruendl}, R.~A., \& {Chu}, Y.-H. 2009,
  \href{http://dx.doi.org/10.1088/0067-0049/184/1/172}{\JournalTitle{\apjs},
  184, 172}

\bibitem[{Harris {et~al.}(2020)Harris, Millman, van~der Walt, Gommers,
  Virtanen, Cournapeau, Wieser, Taylor, Berg, Smith, Kern, Picus, Hoyer, van
  Kerkwijk, Brett, Haldane, del R{'{\i}}o, Wiebe, Peterson,
  G{'{e}}rard-Marchant, Sheppard, Reddy, Weckesser, Abbasi, Gohlke, \&
  Oliphant}]{numpy}
Harris, C.~R., Millman, K.~J., van~der Walt, S.~J., {et~al.} 2020,
  \href{http://dx.doi.org/10.1038/s41586-020-2649-2}{\JournalTitle{Nature},
  585, 357}

\bibitem[{{Henize}(1956)}]{Henize56}
{Henize}, K.~G. 1956,
  \href{http://dx.doi.org/10.1086/190025}{\JournalTitle{\apjs}, 2, 315}

\bibitem[{{Hunter}(2007)}]{matplotlib}
{Hunter}, J.~D. 2007,
  \href{http://dx.doi.org/10.1109/MCSE.2007.55}{\JournalTitle{Computing in
  Science Engineering}, 9, 90}

\bibitem[{{Indebetouw} {et~al.}(2020){Indebetouw}, {Wong}, {Chen}, {Kepley},
  {Lebouteiller}, {Madden}, \& {Oliveira}}]{Indebetouw20}
{Indebetouw}, R., {Wong}, T., {Chen}, C. H.~R., {et~al.} 2020,
  \href{http://dx.doi.org/10.3847/1538-4357/ab5db7}{\JournalTitle{\apj}, 888,
  56}

\bibitem[{{Indebetouw} {et~al.}(2008){Indebetouw}, {Whitney}, {Kawamura},
  {Onishi}, {Meixner}, {Meade}, {Babler}, {Hora}, {Gordon}, {Engelbracht},
  {Block}, \& {Misselt}}]{Indebetouw08}
{Indebetouw}, R., {Whitney}, B.~A., {Kawamura}, A., {et~al.} 2008,
  \href{http://dx.doi.org/10.1088/0004-6256/136/4/1442}{\JournalTitle{\aj},
  136, 1442}

\bibitem[{{Johnson} {et~al.}(2015){Johnson}, {Leroy}, {Indebetouw}, {Brogan},
  {Whitmore}, {Hibbard}, {Sheth}, \& {Evans}}]{Johnson15}
{Johnson}, K.~E., {Leroy}, A.~K., {Indebetouw}, R., {et~al.} 2015,
  \href{http://dx.doi.org/10.1088/0004-637X/806/1/35}{\JournalTitle{\apj}, 806,
  35}

\bibitem[{{Kennicutt}(1998)}]{Kennicutt98}
{Kennicutt}, Robert~C., J. 1998,
  \href{http://dx.doi.org/10.1086/305588}{\JournalTitle{\apj}, 498, 541}

\bibitem[{{Kroupa}(2002)}]{Kroupa02}
{Kroupa}, P. 2002,
  \href{http://dx.doi.org/10.1126/science.1067524}{\JournalTitle{Science}, 295,
  82}

\bibitem[{{Kroupa} {et~al.}(2001){Kroupa}, {Aarseth}, \& {Hurley}}]{Kroupa01}
{Kroupa}, P., {Aarseth}, S., \& {Hurley}, J. 2001,
  \href{http://dx.doi.org/10.1046/j.1365-8711.2001.04050.x}{\JournalTitle{\mnras},
  321, 699}

\bibitem[{{Lada} {et~al.}(2010){Lada}, {Lombardi}, \& {Alves}}]{Lada10}
{Lada}, C.~J., {Lombardi}, M., \& {Alves}, J.~F. 2010,
  \href{http://dx.doi.org/10.1088/0004-637X/724/1/687}{\JournalTitle{\apj},
  724, 687}

\bibitem[{{Larson}(1981)}]{Larson81}
{Larson}, R.~B. 1981,
  \href{http://dx.doi.org/10.1093/mnras/194.4.809}{\JournalTitle{\mnras}, 194,
  809}

\bibitem[{{Longmore} {et~al.}(2013){Longmore}, {Bally}, {Testi}, {Purcell},
  {Walsh}, {Bressert}, {Pestalozzi}, {Molinari}, {Ott}, {Cortese}, {Battersby},
  {Murray}, {Lee}, {Kruijssen}, {Schisano}, \& {Elia}}]{Longmore13}
{Longmore}, S.~N., {Bally}, J., {Testi}, L., {et~al.} 2013,
  \href{http://dx.doi.org/10.1093/mnras/sts376}{\JournalTitle{\mnras}, 429,
  987}

\bibitem[{{Lucchini} {et~al.}(2021){Lucchini}, {D'Onghia}, \&
  {Fox}}]{Lucchini21}
{Lucchini}, S., {D'Onghia}, E., \& {Fox}, A.~J. 2021,
  \href{http://dx.doi.org/10.3847/2041-8213/ac3338}{\JournalTitle{\apjl}, 921,
  L36}

\bibitem[{{McMullin} {et~al.}(2007){McMullin}, {Waters}, {Schiebel}, {Young},
  \& {Golap}}]{McMullin07}
{McMullin}, J.~P., {Waters}, B., {Schiebel}, D., {Young}, W., \& {Golap}, K.
  2007, in Astronomical Society of the Pacific Conference Series, Vol. 376,
  Astronomical Data Analysis Software and Systems XVI, ed. R.~A. {Shaw},
  F.~{Hill}, \& D.~J. {Bell}, 127

\bibitem[{{Meixner} {et~al.}(2006){Meixner}, {Gordon}, {Indebetouw}, {Hora},
  {Whitney}, {Blum}, {Reach}, {Bernard}, {Meade}, {Babler}, {Engelbracht},
  {For}, {Misselt}, {Vijh}, {Leitherer}, {Cohen}, {Churchwell}, {Boulanger},
  {Frogel}, {Fukui}, {Gallagher}, {Gorjian}, {Harris}, {Kelly}, {Kawamura},
  {Kim}, {Latter}, {Madden}, {Markwick-Kemper}, {Mizuno}, {Mizuno}, {Mould},
  {Nota}, {Oey}, {Olsen}, {Onishi}, {Paladini}, {Panagia}, {Perez-Gonzalez},
  {Shibai}, {Sato}, {Smith}, {Staveley-Smith}, {Tielens}, {Ueta}, {van Dyk},
  {Volk}, {Werner}, \& {Zaritsky}}]{Meixner06}
{Meixner}, M., {Gordon}, K.~D., {Indebetouw}, R., {et~al.} 2006,
  \href{http://dx.doi.org/10.1086/508185}{\JournalTitle{\aj}, 132, 2268}

\bibitem[{{Meixner} {et~al.}(2013){Meixner}, {Panuzzo}, {Roman-Duval},
  {Engelbracht}, {Babler}, {Seale}, {Hony}, {Montiel}, {Sauvage}, {Gordon},
  {Misselt}, {Okumura}, {Chanial}, {Beck}, {Bernard}, {Bolatto}, {Bot},
  {Boyer}, {Carlson}, {Clayton}, {Chen}, {Cormier}, {Fukui}, {Galametz},
  {Galliano}, {Hora}, {Hughes}, {Indebetouw}, {Israel}, {Kawamura}, {Kemper},
  {Kim}, {Kwon}, {Lebouteiller}, {Li}, {Long}, {Madden}, {Matsuura}, {Muller},
  {Oliveira}, {Onishi}, {Otsuka}, {Paradis}, {Poglitsch}, {Reach},
  {Robitaille}, {Rubio}, {Sargent}, {Sewi{\l}o}, {Skibba}, {Smith},
  {Srinivasan}, {Tielens}, {van Loon}, \& {Whitney}}]{Meixner13}
{Meixner}, M., {Panuzzo}, P., {Roman-Duval}, J., {et~al.} 2013,
  \href{http://dx.doi.org/10.1088/0004-6256/146/3/62}{\JournalTitle{\aj}, 146,
  62}

\bibitem[{{Miller} {et~al.}(2022){Miller}, {Cioni}, {de Grijs}, {Sun}, {Bell},
  {Choudhury}, {Ivanov}, {Marconi}, {Oliveira}, {Petr-Gotzens}, {Ripepi}, \&
  {van Loon}}]{Miller22}
{Miller}, A.~E., {Cioni}, M.-R.~L., {de Grijs}, R., {et~al.} 2022,
  \href{http://dx.doi.org/10.1093/mnras/stac508}{\JournalTitle{\mnras}},
  \href{http://arxiv.org/abs/2202.09267}{{\sffamily arXiv:2202.09267
  [astro-ph.GA]}}

\bibitem[{{Mizuno} {et~al.}(2001){Mizuno}, {Yamaguchi}, {Mizuno}, {Rubio},
  {Abe}, {Saito}, {Onishi}, {Yonekura}, {Yamaguchi}, {Ogawa}, \&
  {Fukui}}]{Mizuno01}
{Mizuno}, N., {Yamaguchi}, R., {Mizuno}, A., {et~al.} 2001,
  \href{http://dx.doi.org/10.1093/pasj/53.6.971}{\JournalTitle{\pasj}, 53, 971}

\bibitem[{{Nayak} {et~al.}(2018){Nayak}, {Meixner}, {Fukui}, {Tachihara},
  {Onishi}, {Saigo}, {Tokuda}, \& {Harada}}]{Nayak18}
{Nayak}, O., {Meixner}, M., {Fukui}, Y., {et~al.} 2018,
  \href{http://dx.doi.org/10.3847/1538-4357/aaab5f}{\JournalTitle{\apj}, 854,
  154}

\bibitem[{{Nayak} {et~al.}(2016){Nayak}, {Meixner}, {Indebetouw}, {De Marchi},
  {Koekemoer}, {Panagia}, \& {Sabbi}}]{Nayak16}
{Nayak}, O., {Meixner}, M., {Indebetouw}, R., {et~al.} 2016,
  \href{http://dx.doi.org/10.3847/0004-637X/831/1/32}{\JournalTitle{\apj}, 831,
  32}

\bibitem[{{Oka} {et~al.}(2001){Oka}, {Hasegawa}, {Sato}, {Tsuboi}, {Miyazaki},
  \& {Sugimoto}}]{Oka01}
{Oka}, T., {Hasegawa}, T., {Sato}, F., {et~al.} 2001,
  \href{http://dx.doi.org/10.1086/322976}{\JournalTitle{\apj}, 562, 348}

\bibitem[{{Pietrzy{\'n}ski} {et~al.}(2013){Pietrzy{\'n}ski}, {Graczyk},
  {Gieren}, {Thompson}, {Pilecki}, {Udalski}, {Soszy{\'n}ski}, {Koz{\l}owski},
  {Konorski}, {Suchomska}, {Bono}, {Moroni}, {Villanova}, {Nardetto},
  {Bresolin}, {Kudritzki}, {Storm}, {Gallenne}, {Smolec}, {Minniti}, {Kubiak},
  {Szyma{\'n}ski}, {Poleski}, {Wyrzykowski}, {Ulaczyk}, {Pietrukowicz},
  {G{\'o}rski}, \& {Karczmarek}}]{Pietrzynski13}
{Pietrzy{\'n}ski}, G., {Graczyk}, D., {Gieren}, W., {et~al.} 2013,
  \href{http://dx.doi.org/10.1038/nature11878}{\JournalTitle{\nat}, 495, 76}

\bibitem[{{Rahner} {et~al.}(2018){Rahner}, {Pellegrini}, {Glover}, \&
  {Klessen}}]{Rahner18}
{Rahner}, D., {Pellegrini}, E.~W., {Glover}, S. C.~O., \& {Klessen}, R.~S.
  2018, \href{http://dx.doi.org/10.1093/mnrasl/slx149}{\JournalTitle{\mnras},
  473, L11}

\bibitem[{{Rolleston} {et~al.}(2002){Rolleston}, {Trundle}, \&
  {Dufton}}]{Rolleston02}
{Rolleston}, W.~R.~J., {Trundle}, C., \& {Dufton}, P.~L. 2002,
  \href{http://dx.doi.org/10.1051/0004-6361:20021088}{\JournalTitle{\aap}, 396,
  53}

\bibitem[{{Rosolowsky} {et~al.}(2008){Rosolowsky}, {Pineda}, {Kauffmann}, \&
  {Goodman}}]{Rosolowsky08}
{Rosolowsky}, E.~W., {Pineda}, J.~E., {Kauffmann}, J., \& {Goodman}, A.~A.
  2008, \href{http://dx.doi.org/10.1086/587685}{\JournalTitle{\apj}, 679, 1338}

\bibitem[{{Sabbi} {et~al.}(2016){Sabbi}, {Lennon}, {Anderson}, {Cignoni}, {van
  der Marel}, {Zaritsky}, {De Marchi}, {Panagia}, {Gouliermis}, {Grebel},
  {Gallagher}, {Smith}, {Sana}, {Aloisi}, {Tosi}, {Evans}, {Arab}, {Boyer}, {de
  Mink}, {Gordon}, {Koekemoer}, {Larsen}, {Ryon}, \& {Zeidler}}]{Sabbi16}
{Sabbi}, E., {Lennon}, D.~J., {Anderson}, J., {et~al.} 2016,
  \href{http://dx.doi.org/10.3847/0067-0049/222/1/11}{\JournalTitle{\apjs},
  222, 11}

\bibitem[{{S{\'a}nchez} {et~al.}(2007){S{\'a}nchez}, {Alfaro}, \&
  {P{\'e}rez}}]{Sanchez07}
{S{\'a}nchez}, N., {Alfaro}, E.~J., \& {P{\'e}rez}, E. 2007,
  \href{http://dx.doi.org/10.1086/510351}{\JournalTitle{\apj}, 656, 222}

\bibitem[{{Seale} {et~al.}(2012){Seale}, {Looney}, {Wong}, {Ott}, {Klein}, \&
  {Pineda}}]{Seale12}
{Seale}, J.~P., {Looney}, L.~W., {Wong}, T., {et~al.} 2012,
  \href{http://dx.doi.org/10.1088/0004-637X/751/1/42}{\JournalTitle{\apj}, 751,
  42}

\bibitem[{{Seale} {et~al.}(2014){Seale}, {Meixner}, {Sewi{\l}o}, {Babler},
  {Engelbracht}, {Gordon}, {Hony}, {Misselt}, {Montiel}, {Okumura}, {Panuzzo},
  {Roman-Duval}, {Sauvage}, {Boyer}, {Chen}, {Indebetouw}, {Matsuura},
  {Oliveira}, {Srinivasan}, {van Loon}, {Whitney}, \& {Woods}}]{Seale14}
{Seale}, J.~P., {Meixner}, M., {Sewi{\l}o}, M., {et~al.} 2014,
  \href{http://dx.doi.org/10.1088/0004-6256/148/6/124}{\JournalTitle{\aj}, 148,
  124}

\bibitem[{{Sewi{\l}o} {et~al.}(2010){Sewi{\l}o}, {Indebetouw}, {Carlson},
  {Whitney}, {Chen}, {Meixner}, {Robitaille}, {van Loon}, {Oliveira},
  {Churchwell}, {Simon}, {Hony}, {Panuzzo}, {Sauvage}, {Roman-Duval}, {Gordon},
  {Engelbracht}, {Misselt}, {Okumura}, {Beck}, {Hora}, \& {Woods}}]{Sewilo10}
{Sewi{\l}o}, M., {Indebetouw}, R., {Carlson}, L.~R., {et~al.} 2010,
  \href{http://dx.doi.org/10.1051/0004-6361/201014688}{\JournalTitle{\aap},
  518, L73}

\bibitem[{{Sidorin}(2017)}]{Sidorin17}
{Sidorin}, V. 2017, {Quickclump: Identify clumps within a 3D FITS datacube}

\bibitem[{{Solomon} {et~al.}(1987){Solomon}, {Rivolo}, {Barrett}, \&
  {Yahil}}]{Solomon87}
{Solomon}, P.~M., {Rivolo}, A.~R., {Barrett}, J., \& {Yahil}, A. 1987,
  \href{http://dx.doi.org/10.1086/165493}{\JournalTitle{\apj}, 319, 730}

\bibitem[{{Sorai} {et~al.}(2001){Sorai}, {Hasegawa}, {Booth}, {Rubio},
  {Morino}, {Bronfman}, {Handa}, {Hayashi}, {Nyman}, {Oka}, {Sakamoto}, {Seta},
  \& {Usuda}}]{Sorai01}
{Sorai}, K., {Hasegawa}, T., {Booth}, R.~S., {et~al.} 2001,
  \href{http://dx.doi.org/10.1086/320212}{\JournalTitle{\apj}, 551, 794}

\bibitem[{{Sun} {et~al.}(2018){Sun}, {de Grijs}, {Cioni}, {Rubele},
  {Subramanian}, {van Loon}, {Bekki}, {Bell}, {Ivanov}, {Marconi}, {Muraveva},
  {Oliveira}, \& {Ripepi}}]{Sun18}
{Sun}, N.-C., {de Grijs}, R., {Cioni}, M.-R.~L., {et~al.} 2018,
  \href{http://dx.doi.org/10.3847/1538-4357/aabc50}{\JournalTitle{\apj}, 858,
  31}

\bibitem[{{Tokuda} {et~al.}(2019){Tokuda}, {Fukui}, {Harada}, {Saigo},
  {Tachihara}, {Tsuge}, {Inoue}, {Torii}, {Nishimura}, {Zahorecz}, {Nayak},
  {Meixner}, {Minamidani}, {Kawamura}, {Mizuno}, {Indebetouw}, {Sewi{\l}o},
  {Madden}, {Galametz}, {Lebouteiller}, {Chen}, \& {Onishi}}]{Tokuda19}
{Tokuda}, K., {Fukui}, Y., {Harada}, R., {et~al.} 2019,
  \href{http://dx.doi.org/10.3847/1538-4357/ab48ff}{\JournalTitle{\apj}, 886,
  15}

\bibitem[{{Tokuda} {et~al.}(2022){Tokuda}, {Minami}, {Fukui}, {Inoue},
  {Nishioka}, {Tsuge}, {Zahorecz}, {Sano}, {Konishi}, {Chen}, {Sewi{\l}o},
  {Madden}, {Nayak}, {Saigo}, {Nishimura}, {Tanaka}, {Sawada}, {Indebetouw},
  {Tachihara}, {Kawamura}, \& {Onishi}}]{Tokuda22}
{Tokuda}, K., {Minami}, T., {Fukui}, Y., {et~al.} 2022, \JournalTitle{arXiv
  e-prints}, arXiv:2205.00113

\bibitem[{{van der Marel} \& {Kallivayalil}(2014)}]{vanderMarel14}
{van der Marel}, R.~P., \& {Kallivayalil}, N. 2014,
  \href{http://dx.doi.org/10.1088/0004-637X/781/2/121}{\JournalTitle{\apj},
  781, 121}

\bibitem[{Virtanen {et~al.}(2020)Virtanen, Gommers, Oliphant, Haberland, Reddy,
  Cournapeau, Burovski, Peterson, Weckesser, Bright, {van der Walt}, Brett,
  Wilson, Millman, Mayorov, Nelson, Jones, Kern, Larson, Carey, Polat, Feng,
  Moore, {VanderPlas}, Laxalde, Perktold, Cimrman, Henriksen, Quintero, Harris,
  Archibald, Ribeiro, Pedregosa, {van Mulbregt}, \& {SciPy 1.0
  Contributors}}]{scipy}
Virtanen, P., Gommers, R., Oliphant, T.~E., {et~al.} 2020,
  \href{http://dx.doi.org/10.1038/s41592-019-0686-2}{\JournalTitle{Nature
  Methods}, 17, 261}

\bibitem[{{Walborn} {et~al.}(2013){Walborn}, {Barb{\'a}}, \&
  {Sewi{\l}o}}]{Walborn13}
{Walborn}, N.~R., {Barb{\'a}}, R.~H., \& {Sewi{\l}o}, M.~M. 2013,
  \href{http://dx.doi.org/10.1088/0004-6256/145/4/98}{\JournalTitle{\aj}, 145,
  98}

\bibitem[{{Ward} {et~al.}(2016){Ward}, {Oliveira}, {van Loon}, \&
  {Sewi{\l}o}}]{Ward16}
{Ward}, J.~L., {Oliveira}, J.~M., {van Loon}, J.~T., \& {Sewi{\l}o}, M. 2016,
  \href{http://dx.doi.org/10.1093/mnras/stv2424}{\JournalTitle{\mnras}, 455,
  2345}

\bibitem[{{Whitney} {et~al.}(2008){Whitney}, {Sewilo}, {Indebetouw},
  {Robitaille}, {Meixner}, {Gordon}, {Meade}, {Babler}, {Harris}, {Hora},
  {Bracker}, {Povich}, {Churchwell}, {Engelbracht}, {For}, {Block}, {Misselt},
  {Vijh}, {Leitherer}, {Kawamura}, {Blum}, {Cohen}, {Fukui}, {Mizuno},
  {Mizuno}, {Srinivasan}, {Tielens}, {Volk}, {Bernard}, {Boulanger}, {Frogel},
  {Gallagher}, {Gorjian}, {Kelly}, {Latter}, {Madden}, {Kemper}, {Mould},
  {Nota}, {Oey}, {Olsen}, {Onishi}, {Paladini}, {Panagia}, {Perez-Gonzalez},
  {Reach}, {Shibai}, {Sato}, {Smith}, {Staveley-Smith}, {Ueta}, {Van Dyk},
  {Werner}, {Wolff}, \& {Zaritsky}}]{Whitney08}
{Whitney}, B.~A., {Sewilo}, M., {Indebetouw}, R., {et~al.} 2008,
  \href{http://dx.doi.org/10.1088/0004-6256/136/1/18}{\JournalTitle{\aj}, 136,
  18}

\bibitem[{{Wong} {et~al.}(2019){Wong}, {Hughes}, {Tokuda}, {Indebetouw},
  {Onishi}, {Band urski}, {Chen}, {Fukui}, {Glover}, {Klessen}, {Pineda},
  {Roman-Duval}, {Sewi{\l}o}, {Wojciechowski}, \& {Zahorecz}}]{Wong19}
{Wong}, T., {Hughes}, A., {Tokuda}, K., {et~al.} 2019,
  \href{http://dx.doi.org/10.3847/1538-4357/ab46ba}{\JournalTitle{\apj}, 885,
  50}

\bibitem[{{Wong} {et~al.}(2022){Wong}, {Oudshoorn}, {Sofovich}, {Green},
  {Shah}, {Indebetouw}, {Meixner}, {Hacar}, {Nayak}, {Tokuda}, {Bolatto},
  {Chevance}, {De Marchi}, {Fukui}, {Hirschauer}, {Jameson}, {Kalari},
  {Lebouteiller}, {Looney}, {Madden}, {Onishi}, {Roman-Duval}, {Rubio}, \&
  {Tielens}}]{Wong22}
{Wong}, T., {Oudshoorn}, L., {Sofovich}, E., {et~al.} 2022, \JournalTitle{arXiv
  e-prints}, arXiv:2206.06528

\bibitem[{{Yamaguchi} {et~al.}(2001){Yamaguchi}, {Mizuno}, {Mizuno}, {Rubio},
  {Abe}, {Saito}, {Moriguchi}, {Matsunaga}, {Onishi}, {Yonekura}, \&
  {Fukui}}]{Yamaguchi01}
{Yamaguchi}, R., {Mizuno}, N., {Mizuno}, A., {et~al.} 2001,
  \href{http://dx.doi.org/10.1093/pasj/53.6.985}{\JournalTitle{\pasj}, 53, 985}

\bibitem[{{Zivick} {et~al.}(2019){Zivick}, {Kallivayalil}, {Besla}, {Sohn},
  {van der Marel}, {del Pino}, {Linden}, {Fritz}, \& {Anderson}}]{Zivick19}
{Zivick}, P., {Kallivayalil}, N., {Besla}, G., {et~al.} 2019,
  \href{http://dx.doi.org/10.3847/1538-4357/ab0554}{\JournalTitle{\apj}, 874,
  78}

\end{thebibliography}
    
\end{document}